\documentclass[conference]{IEEEtran}
\pdfoutput=1

\usepackage{cite}
\usepackage{amsmath,amssymb,amsfonts}
\usepackage{epsfig, subfigure, amsmath, amssymb}
\usepackage{algorithm}
\usepackage{algpseudocode}
\usepackage{graphicx}
\usepackage{tikz}
\usetikzlibrary{shapes}
\usepackage{textcomp}
\usepackage{xcolor}
\usepackage{multirow}
\usepackage{makecell}
\def\BibTeX{{\rm B\kern-.05em{\sc i\kern-.025em b}\kern-.08em
    T\kern-.1667em\lower.7ex\hbox{E}\kern-.125emX}}

\usepackage{optidef}
\usepackage[normalem]{ulem}
\usepackage{adjustbox}
\usepackage{xcolor}
\usepackage{paralist}
\usepackage{fancyhdr}
\usepackage{url}

\usepackage{soul}
\fancypagestyle{mypagestyle}{
  \fancyhf{} 
  \fancyfoot[C]{Approved for Public Release on 12 Mar 2024. Distribution Is Unlimited. Case Number: 2024-0109  (Original Case Number(s):  AFRL-2024-0499)}
  
}
\begin{document}

\title{SHEATH: Defending Horizontal Collaboration for Distributed CNNs against Adversarial Noise}

\author{
    Muneeba Asif\textsuperscript{*},
    Mohammad Kumail Kazmi\textsuperscript{*},
    Mohammad Ashiqur Rahman\textsuperscript{*},
    Syed Rafay Hasan\textsuperscript{\dag}, Soamar Homsi\textsuperscript{\ddag}\\
    
    \textsuperscript{*}Department of Electrical and Computer Engineering, Florida International University, USA\\
    \textsuperscript{\dag}Department of Computer Science, Tennessee Technological University, USA\\    
    \textsuperscript{\ddag}Information Warfare Division, AirForce Research Laboratory, USA\\
    
    \{masif004, mkazm004, marahman\}@fiu.edu, shasan@tntech.edu, soamar.homsi@us.af.mil
}




\maketitle
\maketitle
\thispagestyle{mypagestyle} 
\pagestyle{mypagestyle}     

\begin{abstract}
\label{sec:abstract}
As edge computing and the Internet of Things (IoT) expand, horizontal collaboration (HC) emerges as a distributed data processing solution for resource-constrained devices. In particular, a convolutional neural network (CNN) model can be deployed on multiple IoT devices, allowing distributed inference execution for image recognition while ensuring model and data privacy. Yet, this distributed architecture remains vulnerable to adversaries who want to make subtle alterations that impact the model, even if they lack access to the entire model. Such vulnerabilities can have severe implications for various sectors, including healthcare, military, and autonomous systems. However, security solutions for these vulnerabilities have not been explored. This paper presents a novel framework for Secure Horizontal Edge with Adversarial Threat Handling (SHEATH) to 
detect adversarial noise and eliminate its effect on CNN inference by recovering the original feature maps.
Specifically, SHEATH aims to address vulnerabilities without requiring complete knowledge of the CNN model in HC edge architectures based on sequential partitioning. It ensures data and model integrity, offering security against adversarial attacks in diverse HC environments. Our evaluations demonstrate SHEATH's adaptability and effectiveness across diverse CNN configurations.

\end{abstract}


\section{Introduction}
\label{sec:intro}


The internet of things (IoT) infrastructures can be integrated with artificial intelligence (AI) and machine
learning (ML) to create artificial intelligence of things (AIoT). This advancement enhances urban living and sustainability in
the form of developing smart cities, which address real-world
problems like smart parking~\cite{reuters2023}, fostering next-generation smart grid solutions~\cite{esenogho2022integrating}, and enabling large-scale sensor deployments~\cite{seng2022artificial}. 
To enhance the efficiency of IoT operations using AIoT, there must be a focus on data management and analysis: collecting, delivering, and processing data, making decisions, and executing them
accordingly~\cite{dataconomy2022}.
These tasks require high computing power, leading IoT systems to leverage cloud platforms and shift computation to edge nodes for efficiency. However, IoT devices often have infrastructure limitations or failures that prevent cloud access.
Moreover, communication with the cloud creates a
single point of failure, and these
channels are cyberattack-prone. Cloud reliance also raises privacy concerns, given the third-party access to information. 

To address these security challenges, researchers have proposed distributing these tasks on edge devices or servers; this comes with additional benefits, including latency reduction, privacy preservation, and energy efficiency~\cite{maomodnn, neurosurgeon}. However, these edge IoT devices are often resource-constrained, such as low processing power, limited memory, and energy~\cite{pereira2020challenges}. These limited processing capabilities highlight the need for external computational support. 
This support is offered by horizontal collaboration (HC), which enables cooperation among edge nodes 
in decentralized networks, thereby improving performance by sharing computational tasks~\cite{shi2016edge, wangsurvey}. 
For example, the inference of trained convolutional neural networks (CNNs) can be partitioned onto multiple edge devices. This collaborative inference can be achieved by (i) sequentially dividing the model across multiple devices~\cite{li2019edge} or (ii) segmenting data for model parallelism followed by result amalgamation~\cite{zeng2021coedge}. 

\begin{figure*}[!t]
    \centering
    \includegraphics[width=0.95\textwidth, height=4.5cm]{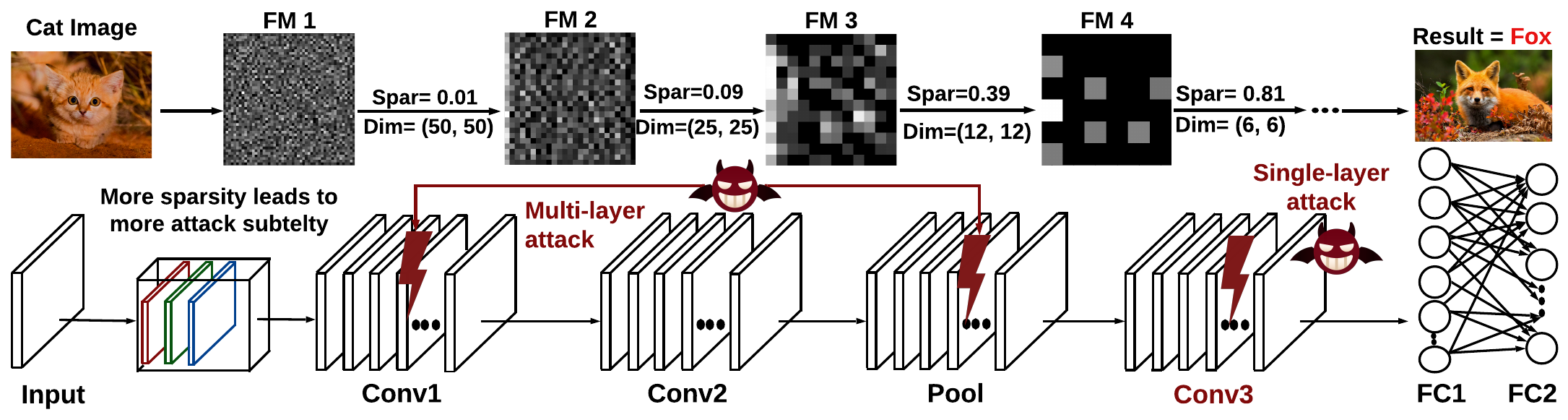}
    \caption{\small Horizontal CNN partitioning across IoT devices. Adversarial noise at \textit{``Conv3"} results in a \textit{``fox"} misclassification of a \textit{``cat"} image. The adversary can launch multi-node (i.e., multi-layer) attacks, as shown on the \textit{``Conv1"} and \textit{``Pool1"} layers. Increased sparsity in deeper CNN layers intensifies the subtlety of these adversarial attacks, which makes them harder to detect.}   
    \label{fig:hcsetup}
    \vspace{-5pt}
\end{figure*}

While this collaboration allows IoT devices/edge nodes to execute expensive ML inference operations, it provides several essential security benefits. For instance, each device possesses only parts of the data or model, limiting the exposure of sensitive information~\cite{zewe2022}. Such privacy is essential in areas like the military, healthcare, autonomous driving, and smart homes, where timely and accurate responses are essential. Misclassifications in these domains can lead to severe consequences, such as unintended casualties, wrong treatments, accidents, or security breaches. 
However, HC edge environments are still vulnerable to cyberattacks as nodes are often untrusted. Adversaries can compromise one or more nodes and inject noise into the feature maps at these nodes, thus compromising model integrity even without full model access, as shown in Fig.~\ref{fig:hcsetup}. 
Such a noise-based inference attack on capsule networks in an HC environment was demonstrated by Adeymo et al., causing an average accuracy drop of 62\%~\cite{adeyemo2021security}.

To preserve the data and model integrity in the HC environment, the edge devices must be resilient against cyberattacks. Most existing solutions for detecting and defending against adversarial attacks in the HC environment require full knowledge of the CNN~\cite{khalid2019fademl} or involve retraining the model for increased robustness~\cite{papernot2016distillation, lee2020gradient}. However, these methods mainly apply to the segmented/amalgamated approach and not to sequential partitioning; hence, they fail to consider the dynamic aspects of HC. This is especially problematic for dynamic use cases, e.g., when the nodes are a fleet of drones. Therefore, we propose \textbf{S}ecure \textbf{H}orizontal \textbf{E}dge with \textbf{A}dversarial \textbf{T}hreat \textbf{H}andling (SHEATH), a novel framework to address the aforementioned vulnerabilities in sequential partitioning-based HC edge architectures without requiring full knowledge of the CNN model. 

As mentioned earlier, the different layers of the HC-based CNN model are deployed on different nodes that are often untrusted. Some of these nodes can be malicious or compromised where attacks occur, i.e., adversarial noise is added. SHEATH is a lightweight technique that defends the inference against such attacks on a selected untrusted node. SHEATH itself will be deployed on a trusted node to ensure the process. This node can be semi-honest, i.e., it may be curious about the input data or the inference result but will execute the SHEATH operation correctly. Since SHEATH acts on an intermediate layer of the CNN model, it reveals no useful information.
%
SHEATH comprises two modules: (i) \textit{Detect} and (ii) \textit{Recover}. The \textit{Detect} module further consists of two parts \textit{PseudoNet} and the \textit{Comparator}. 
\textit{PseudoNet} computes the expected feature maps of the untrusted node to serve as a basis of the non-noisy (reference) feature maps for the \textit{Comparator}. \textit{PseudoNet} is a less computational-intensive copy of the node that SHEATH targets to secure. For example, if the target node is a convolutional layer with 64 kernels, its corresponding \textit{PseudoNet} will be a convolutional layer but with fewer kernels. The goal is to achieve a trade-off between accurate noise detection and heavy computational load. 
After this, the \textit{Comparator} evaluates the difference between the outputs of both \textit{PseudoNet} and 
the feature maps from the untrusted node. If this difference exceeds the predefined threshold, i.e., the adversarial noise is detected, the \textit{Recover} module is activated. 
\textit{Recover} infers the correct/expected feature maps and forwards them to the next node in the inference chain.
The design of SHEATH ensures the integrity 
of the CNN model deployed on HC edge devices with an acceptable overhead and seamless integration.

However, SHEATH's deployment strategy can vary based on factors such as system architecture, computational capabilities, network bandwidth, security requirements, and specific use-case demands. We discuss multiple deployment scenarios 
and evaluate the framework on various CNN models and datasets. Overall, our contributions are threefold as follows:
\begin{itemize}
\item We conduct an in-depth analysis of the associated vulnerabilities and the implications of single and multi-layer adversarial noise attacks on CNN models deployed on HC edge devices.
\item We propose SHEATH, a novel framework to detect adversarial noise and eliminate the effects of the aforementioned attacks on the HC-based CNNs by recovering the original feature maps. We also discuss various deployment strategies for SHEATH.
\item We comprehensively evaluate SHEATH's effectiveness across various CNN models and datasets, demonstrating its adaptability and robustness in diverse AIoT scenarios.
\end{itemize}

The rest of the paper is organized as follows: Section~\ref{sec:background} elucidates edge computing in HC, potential attacks and provides the research motivation. The threat model is discussed in Section~\ref{sec:threat_model}. Section~\ref{sec:case_study} presents the case studies, Section~\ref{sec:problem_formulation} presents the problem formulation, and Section~\ref{sec:proposed_solution} discusses the proposed framework along with different deployment scenarios. Performance is empirically validated in Section~\ref{sec:evaluation}. Pertinent literature is reviewed in Section~\ref{sec:related_works}; finally, the paper is concluded in Section~\ref{sec:conclusion}.

\section{Theoretical Foundations}
\label{sec:background}

This section overviews the distributed CNNs in an HC setup, their significance, their vulnerability to adversarial attacks, and finally, highlights the motivation behind our research.

\subsection{HC Edge Environment for Distributed CNNs}

HC refers to the interaction between devices or nodes within a decentralized network. Contrary to vertical collaboration, where data is relayed from edge devices to centralized systems and back, HC utilizes multiple devices to share tasks and resources. This type of collaboration is substantial in edge computing and IoT settings where devices are resource-constrained. In traditional centralized models, a central node/server delegates the tasks to peripheral nodes. However, HC bypasses this central entity, allowing edge devices to interact directly with each other. By distributing tasks and leveraging the collective computational power of multiple devices, HC enhances system resilience and throughput, reduces latency, and optimizes resource utilization. Teixeira et al. highlighted HC's advantages in IoT settings, like reduced delays and increased privacy~\cite{teixeira2017horizontal}. Kumar et al. introduced a framework where devices collaboratively decide task distribution using a consensus method~\cite{kumar2019framework}. 
Additionally, HC principles can be applied to federated learning, as demonstrated by Yang et al., where devices work together for model training without sharing raw data, ensuring data privacy~\cite{yang2019federated}.

HC distributes CNN models across edge devices by partitioning and deploying them in ways such as layer-wise, feature-wise, or data-wise, facilitating collaborative inference despite the limited resources of each device~\cite{wang2019cnn_partitioning}. Layer-wise partitioning has each device process specific layers of a CNN and then pass the results to the next device for further computation. Some of the advantages of HC are (i) \textit{efficiency and speed}, (ii) \textit{data privacy and security}, and (iii) \textit{reduced bandwidth consumption} as there is minimized data transmission due to edge processing. However, distributed CNNs in HC remain vulnerable to adversarial attacks.

\subsection{Node Collaboration and Security in Edge Computing}

To understand the problem practically, consider an image classification CNN model.  It requires significant computational power and faces privacy issues with cloud computing. Edge computing offers a solution, but IoT devices at the edge have limited resources. Hence, HC, a method in which multiple edge nodes cooperate, is used to mitigate their resource constraints in data processing. In this collaboration, the CNN model is initially trained offline, and the inference is distributed among the multiple collaborating nodes. This means that the CNN model's various \textbf{layers} will be \textbf{distributed on different nodes}. Each node will process its share of the model and forward the output to the next node. This method is computationally efficient, but nodes are chosen dynamically based on availability, leading to \textbf{node trust concerns}, meaning there will be some trusted and untrusted nodes. Trusted nodes are semi-honest, i.e., they will perform their share of model computation correctly without being vulnerable to compromise. Contrarily, \textbf{an untrusted node may not necessarily be malicious}. It may be either benign or malicious, but it is not known. Hence, the untrusted node may be vulnerable to compromise, and any attacks in these nodes must be detected, and the original node outputs recovered thereupon. Thus, SHEATH, the proposed defense framework, \textbf{will be deployed on a trusted node} to protect (detect and recover) a designated untrusted node. Upon identifying an untrusted node as malicious, removing it for future operations may be considered a possible defense. However, SHEATH aims to correct detected noise rather than node removal, allowing for a resilient system that maintains operational integrity even when nodes behave unpredictably, based on the premise that nodes may not be inherently malicious and can return to a trusted state. The correction process is carried out by nodes equipped with pseudoNet, integrating seamlessly with the machine learning framework to ensure minimal disruption and maintain data flow continuity for real-time operations. 

\subsection{Adversarial Attacks in HC-based Distributed CNNs}

 Distributing CNN models over HC-based edge devices is a decentralized approach to enhance computational efficiency and collaboration among multiple devices. While devices in such a network might access only parts of the data or the model, it would be naive to assume that this setup is resilient against cyberattacks. Despite limited access, adversaries can introduce noise to the intermediate data, compromising the overall model integrity. These manipulations, albeit subtle, can have a cascading impact on the model, leading to significant errors or misinterpretations. The sequential nature of the HC setup may further delay the detection of these threats. This can have notable ramifications, especially in healthcare, defense, or autonomous navigation.

To launch an attack on an intermediary node in the model, adversaries can target the statistical properties of that node's feature maps (FMs). Introducing minimal but meaningful noise can degrade the model's efficacy without getting detected. An illustrative example is demonstrated in Fig.~\ref{fig:hcsetup}, where a CNN model, distributed across several IoT devices, processes an image of a cat. Each device processes different model layers, and the feature maps become more sparse as the data transitions among devices. However, an attack on the \textit{``Conv3"} layer, though initially unnoticeable, causes the model to mislabel the cat as a fox. Attackers can also target multiple layers, as shown in \textit{``Conv1"} and \textit{``Pool1"}, further amplifying the attack's impact. As the FMs advance through the model, noise is embedded and increasingly concealed by decreasing feature densities, making its detection harder. Hence, a robust security framework is required for attack resilience in HC-based CNNs distributed on edge devices.

\subsection{Research Motivation}

With the growing adoption of CNNs in decentralized systems, areas such as healthcare, military, and autonomous driving are witnessing innovative data processing. While HC enhances efficiency, it also exposes systems to potential security risks. Despite limited model knowledge, adversaries can inject noise into one or more intermediary nodes, causing misinterpretations. For example, an erroneous image interpretation in military contexts could cause unintended harm; in healthcare, it might cause incorrect patient treatment; for autonomous vehicles, a simple error reading a traffic sign can be dangerous to road safety. While the distributed nature of CNNs in HC architectures offers advantages in terms of latency reduction and privacy, it does not guarantee immunity from adversarial interventions. Studies, including those by Adeymo et al., demonstrate how even partial model access can be exploited by adversaries~\cite{adeymo2023stain}. It must be noted that adversarial training is widely recognized as a key defense mechanism against such attacks in the literature, with numerous studies advocating its effectiveness~\cite{mohammadi2019end, shafahi2019adversarial, miller2020adversarial}. However, this method requires model retraining, which is not applicable to HC-based distributed CNN models due to the unpredictable nature of attack locations within the model. Consequently, there's a critical need for innovative solutions that can preemptively identify and mitigate adversarial noise without the exhaustive requirement of model retraining, especially in environments where the specific segments under attack are not predefined. Motivated by these challenges, our research aims to develop a robust and efficient framework to detect potential adversarial intrusions and provide recovery strategies for horizontally collaborating edge nodes that are running parts of the CNN model. The goal is the integrity preservation of CNN models in HC environments while optimizing the tradeoff between security and computational performance.

\section{Threat Model}
\label{sec:threat_model}

This section overviews our assumptions and the adversary's knowledge and attack goals and summarises attack techniques.

\subsection{Attack Assumptions}

We consider the following attack assumptions for our work.

\begin{itemize}
    \item Communication channels integrity: 
    Data transmission from one node to another is secure. Therefore, the attack can occur only at the malicious node. 
    
    \item Node integrity: Some nodes in the HC architecture are trusted (can be semi-honest), while the rest are untrusted (can be malicious or compromised). Trusted nodes are essential for SHEATH to be deployed and executed.
    
    \item Limited access: 
    A malicious node can only add noise to the feature maps by accessing the input, output, and parameters of the CNN layers implemented \textit{at this node.}   
    
    \item Attack types: One node deploys one layer of the HC-based CNN model. Thus, attackers may perform single-layer attacks by attacking a single node or a multi-layer attack by attacking multiple, non-consecutive nodes with varying levels of noise.
\end{itemize}

\subsection{Attack Goal} 

The attacker's primary goal is to discreetly disrupt the system functionality by injecting noise into one or more intermediate nodes. By adding noise to one of the layers deployed on one of the multiple edge devices, the attacker aims to weaken the model's decision process and make it more vulnerable, leading to erroneous inferences.


\subsection{Adversarial Capabilities}

We assume the attacker's capabilities include the following:

\begin{itemize}
\item {Partial model access:} Attackers can access and potentially modify certain parts of the model, specifically those in the malicious node(s).
\item {Data manipulation:} Attackers can introduce noise or other perturbations into the data or intermediate feature maps that flow through the malicious node(s).
\item {Stealthy intervention:} Attackers can make their modifications subtle, i.e., hard to detect. They can leverage the statistical characteristics of the intermediary node's feature maps to make the attack stealthy.
\end{itemize}

\subsection{Attack Technique}
 In a distributed or collaborative learning environment, certain nodes or parties might only have access to specific segments of a model. However, this limited access does not preclude the potential for malicious activity. Leveraging this partial access, an attacker can initiate statistical attacks to compromise the model's integrity. By subtly altering the statistical properties of the feature maps, the attacker can induce deviations in the model's behavior. Such alterations might include modifying parameter values while ensuring that their aggregate statistical characteristics, like mean and variance, remain unchanged. This stealthy approach makes the perturbations challenging to detect, especially since the overall statistical footprint of the parameters appears untouched. The attacker's goal, in these scenarios, often revolves around introducing subtle biases, inducing misclassifications, or degrading the overall performance of the model.

Given the complexity of modern deep learning models, even a minor alteration in one segment can trigger cascading effects across the model. This can fulfill the attacker's objectives while remaining undetected. Given a set of model parameters \( P \) with mean \( \mu \) and variance \( \sigma^2 \), an attacker can alter this set to produce a new set \( P' \) such that \( \mu(P') = \mu(P) \) and \( \sigma^2(P') = \sigma^2(P) \). The altered parameters \( P' \) are generated as:
\begin{equation}
\label{eq:attack_general}
    P' = f(P, \epsilon)
\end{equation}
Here, \( f \) is a transformation function representing the statistical attack. This function takes the original parameters and a perturbation vector \( \epsilon \) to produce the altered parameters. 
This vector is designed so that it doesn't drastically change the mean and variance of the original parameters.

\subsection{Attack Vector}

In an HC-edge environment, we consider a set of edge nodes represented by $E = \{ e_1, e_2, \dots, e_n \}$, where $n$ is the total number of edge nodes present in the system. Furthermore, consider a CNN whose layers are denoted by the set $L = \{ l_1, l_2, \dots, l_m \}$, with $m$ being the number of layers in the CNN. The core part of this attack vector is the 
noise injection into select layers of the CNN distributed across the edge nodes. We represent this noise matrix by $N$, wherein each element $N_{i,j}$ represents the noise introduced at layer $l_i$ for edge node $e_j$. Mathematically, $N$ is defined as:
\[ N = \begin{bmatrix} N_{1,1} & N_{1,2} & \dots & N_{1,n} \\ N_{2,1} & N_{2,2} & \dots & N_{2,n} \\ \vdots & \vdots & \ddots & \vdots \\ N_{m,1} & N_{m,2} & \dots & N_{m,n} \end{bmatrix} \]

Given a data sample $d$, the genuine output of the model without noise interference at layer $l_i$ in edge node $e_j$ is denoted by $M(d, l_i, e_j)$. However, the output becomes perturbed under the influence of the attack vector. This perturbed output is represented by $M'(d, l_i, e_j)$ and is given by:
\begin{equation}
\label{eq:noise_injection}
    M'(d, l_i, e_j) = M(d, l_i, e_j) + N_{i,j}
\end{equation}
Upon successful noise injection, the malicious node subtly blends with the regular operations of the network while evading detection. This highlights the need for a robust defense framework for HC architectures with partitioned CNN models.




\section{Case Study and Observations}
\label{sec:case_study}

We devised a custom CNN model to understand the potential effect of an adversary operating on HC-based CNN. 

\begin{table*}[!htbp]
    \caption{Variation in the Statistical Mean of the Feature Map against varying Percentage of Noisy Nodes}
    \label{tab:stealth-analysis}
    \centering
    \begin{tabular}{|c|c|c|c|c|c|c|c|c|c|c|c|c|}
    \hline
        CNN Model & Datasets & Mean Before Noise & 5\% & 10\% & 15\% & 20\% & 25\% & 30\% & 35\% & 40\% & 45\% & 50\% \\ \hline
        EdgeCNN & MNIST & 0.8668 & 0.9109 & 0.9511 & 0.9880 & 1.0535 & 1.0764 & 1.1200 & 1.1539 & 1.2194 & 1.2559 & 1.2954 \\ \hline
        LeNet-5 & MNIST & 1.0716 & 1.1198 & 1.1687 & 1.2236 & 1.2833 & 1.3342 & 1.3770 & 1.4453 & 1.4949 & 1.5472 & 1.6061 \\ \hline
        LeNet-5 & Fashion & 1.1062 & 1.1622 & 1.2217 & 1.2611 & 1.3537 & 1.3710 & 1.4208 & 1.4561 & 1.5646 & 1.5957 & 1.6559 \\ \hline
    \end{tabular}
\end{table*}

\begin{table*}[!htbp]
    \caption{Variation in the Statistical Standard Deviation of the Feature Map against varying Percentage of Noisy Nodes}
    \label{tab:stealth-analysis2}
    \centering
    \begin{tabular}{|c|c|c|c|c|c|c|c|c|c|c|c|c|}
    \hline
        CNN Model & Datasets & Stdev Before Noise & 5\% & 10\% & 15\% & 20\% & 25\% & 30\% & 35\% & 40\% & 45\% & 50\% \\ \hline
        EdgeCNN & MNIST & 1.4198 & 1.4284 & 1.4524 & 1.4508 & 1.4649 & 1.4746 & 1.4743 & 1.4690 & 1.4883 & 1.4808 & 1.4902 \\ \hline
        LeNet-5 & MNIST & 1.5921 & 1.6141 & 1.6128 & 1.6425 & 1.6590 & 1.6412 & 1.6551 & 1.6659 & 1.6785 & 1.6694 & 1.6714 \\ \hline
        LeNet-5 & Fashion & 1.2264 & 1.2381 & 1.2722 & 1.2847 & 1.3100 & 1.3070 & 1.3240 & 1.3371 & 1.3428 & 1.3681 & 1.3543 \\ \hline
    \end{tabular}
    \vspace{-6pt}
\end{table*}

\subsection{System and Model Description}

We consider an HC system that has partitioned the CNN model onto multiple edge devices/nodes, as shown in Fig.~\ref{fig:hcsetup}. We designed a reference CNN model named \textit{EdgeCNN} to analyze and classify the MNIST dataset, a collection of handwritten digits commonly used for image classification tasks. This model has an architecture of two convolutional layers (with 32 filters), followed by a max-pooling layer, another convolutional layer (64 filters), and finally, two fully connected layers. The network is trained using a Stochastic Gradient Descent (SGD) optimizer with a learning rate of 0.001. The first node receives the input image and extracts the initial features. The subsequent nodes run \textit{``Conv1", ``Conv2", ``Pool1"}, and \textit{``Conv3"}, respectively, which are the convolutional and pooling layers. We assume that the node running the third convolutional layer is malicious. Hence, noise is introduced in the feature maps obtained from \textit{``Conv3"}. Finally, the following nodes process the fully connected (FC) layers, \textit{``FC1"} and  \textit{``FC2"}, to yield the model's output. 

\begin{figure}[!t]
\centering
        \subfigure[\label{fig:noise-acc}]
        {
        \includegraphics[width=0.45\columnwidth, height=3.3cm]{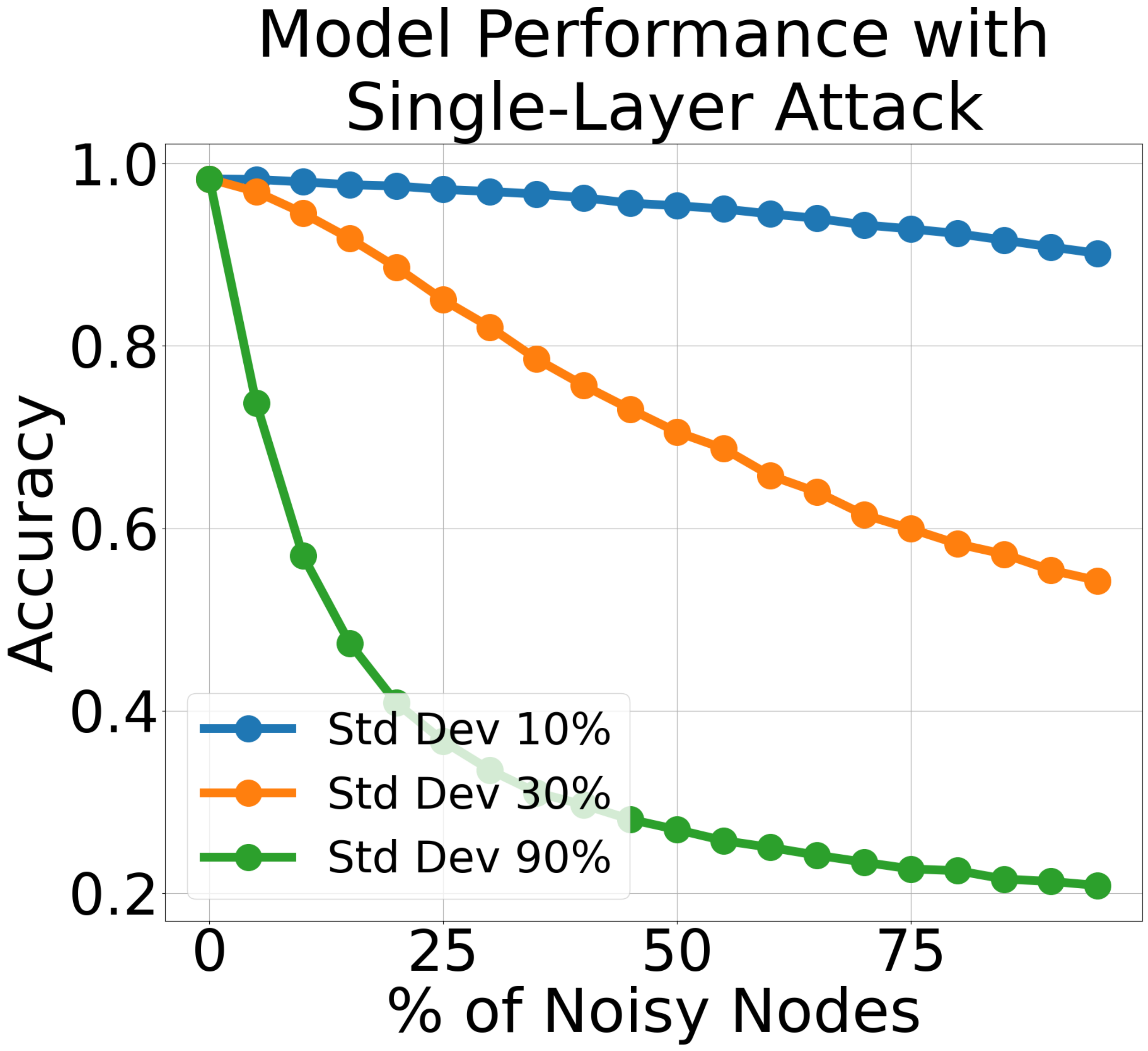}
        }
        \subfigure[\label{fig:mse-acc}]
        {
        \includegraphics[width=0.45\columnwidth, height=3.3cm]{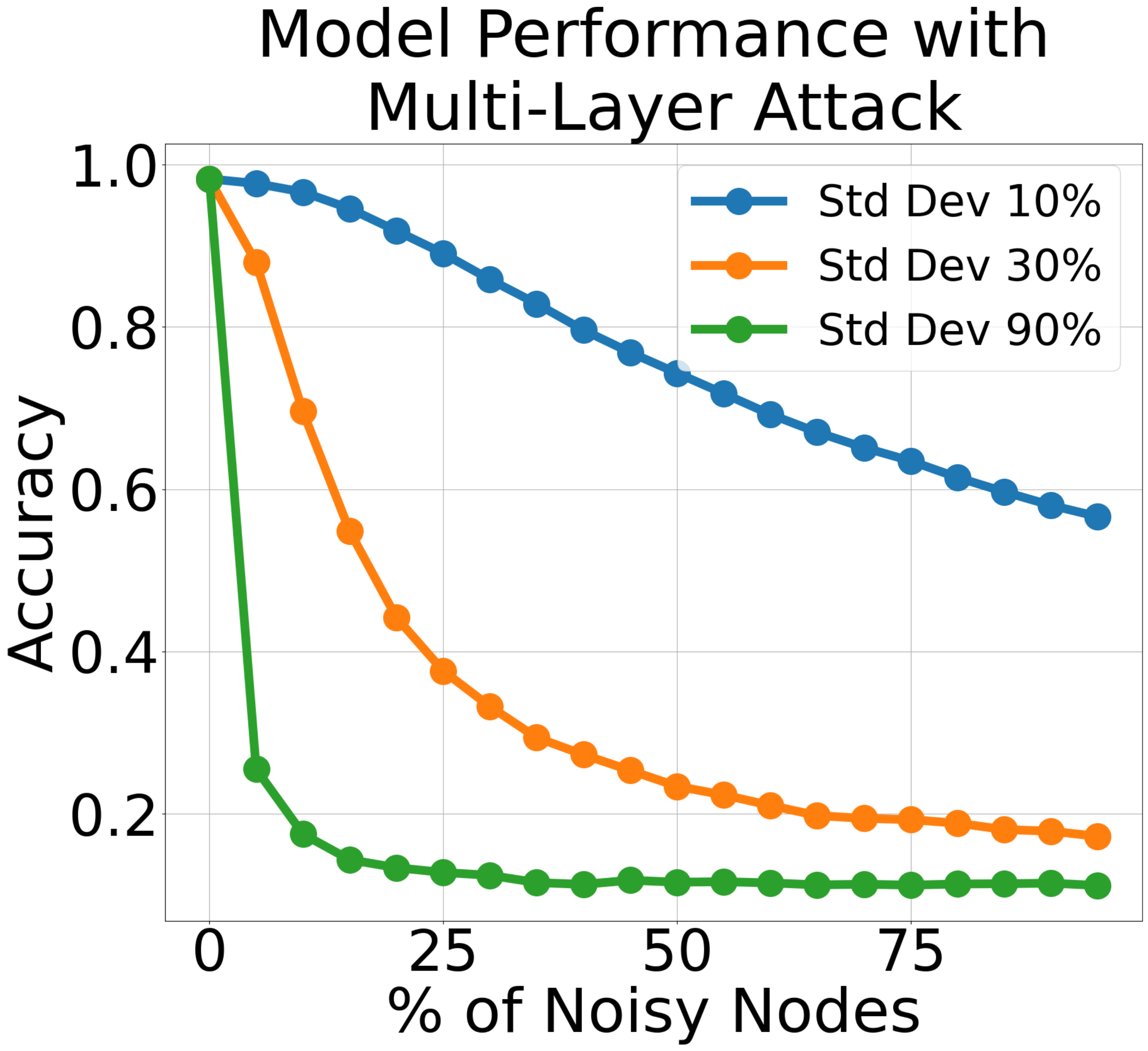}
        }
\vspace{-6pt}
\caption{\small Model Accuracy vs. Noise Parameters in Feature Vector when (a) noise is injected in the third convolutional layer's feature maps and (b) noise is injected in two non-consecutive convolutional layers \textit{``Conv1"} and \textit{``Conv3"}. }
\label{fig:case-study}
\vspace{-6pt}
\end{figure}

\begin{figure}[!t]

        \subfigure[\label{fig:edge}]
        {
        \includegraphics[width=0.46\columnwidth]{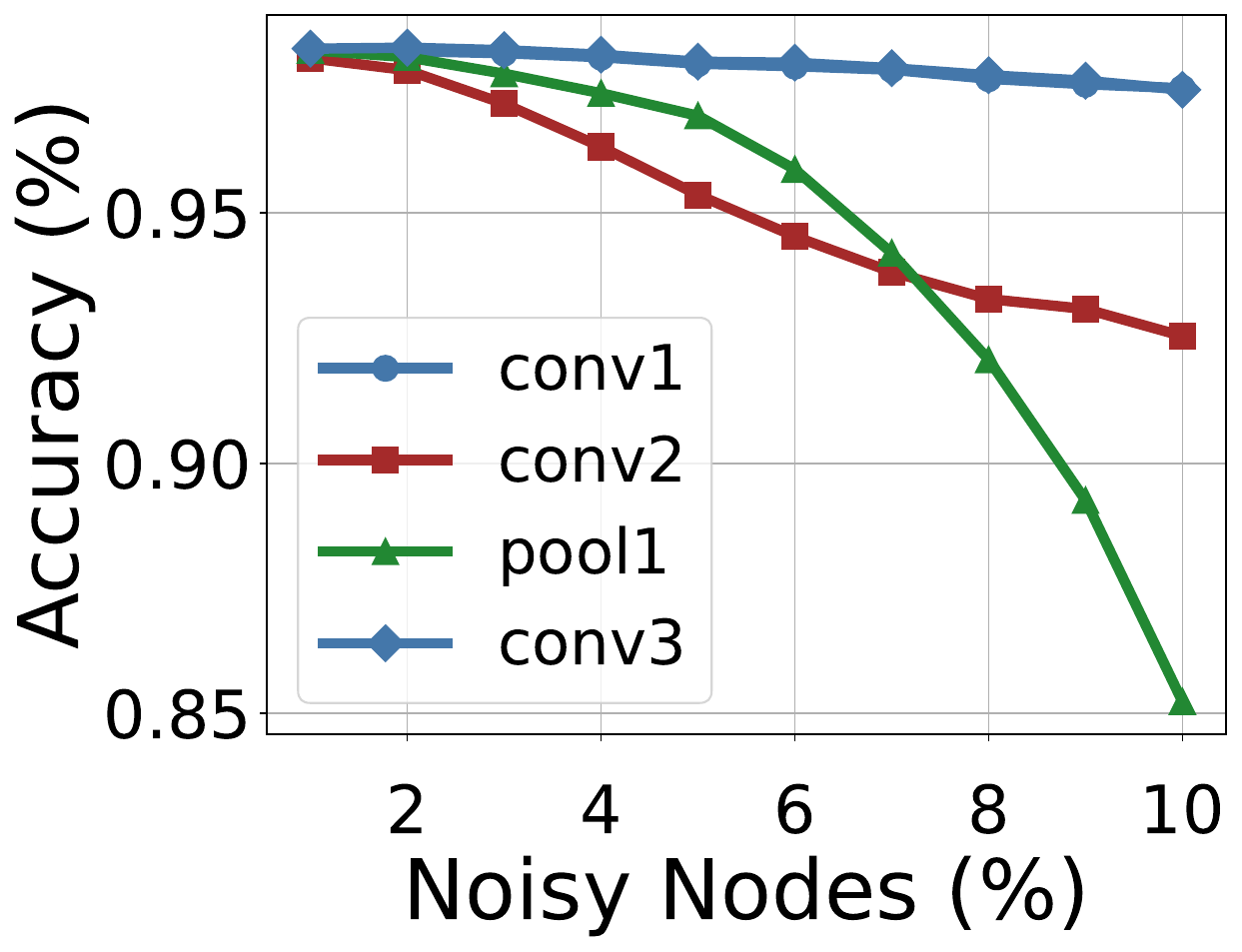}
        }
        \subfigure[\label{fig:lenet}]
        {
        \includegraphics[width=0.46\columnwidth]{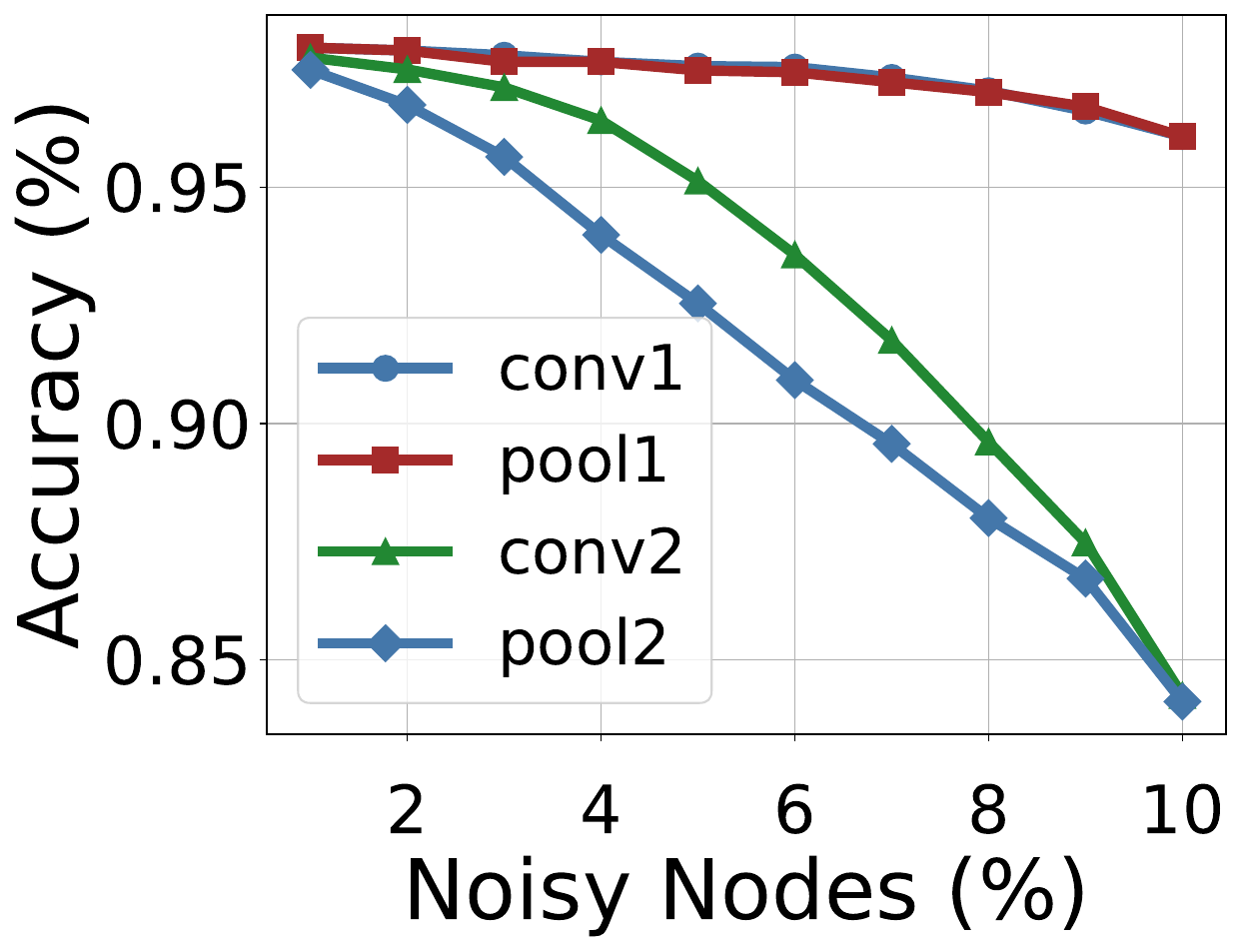}
        }
\vspace{-6pt}
\caption{\small Impact of noise injection in different layers of (a) EdgeCNN and (b) LeNet CNN architectures in an HC setup.}
\label{fig:layer}
\vspace{-12pt}
\end{figure}

\subsection{Noise Injection}
We implemented a custom layer introducing Gaussian noise into the inputs to test the model's performance. This Gaussian noise is defined in Eq.~\ref{eq:noise}. Here $I$ is the input, $\mu$ is the mean of the input, $\sigma$ is the standard deviation of the input, $G$ is the Gaussian distribution, $B$ is the Bernoulli distribution, $np$ is the noise percentage which is a binary mask dictating where the noise is to be added, and $sp$ is the standard deviation percentage (noise magnitude) that was integrated into the EdgeCNN model. The model's performance was evaluated on the MNIST test set. This layer assessed the EdgeCNN model's robustness to various noise levels and impacted nodes.
\begin{equation}
\label{eq:noise}
I' = I + G(\mu, sp*\sigma) \cdot B(np)
\end{equation}
%

\subsection{Observations}

We appended a noise layer based on Eq.~\ref{eq:noise} after the \textit{``Conv3"} layer to assess the effects of varying the percentage of nodes impacted by noise ($np$) and the noise magnitude ($sp$). On the MNIST test set, the model's accuracy was determined for each $np$ and $sp$ combination. This test yielded accuracy variation with increasing noisy nodes across different magnitudes. Key observations include: (i) the \textit{EdgeCNN} model initially attained a 98.87\% accuracy without noise, (ii) the model's performance deteriorates with noise, contingent on the percentage of noisy nodes and its magnitude, and (iii) Fig.~\ref{fig:case-study} demonstrates that even minor noise can considerably reduce accuracy, particularly when the magnitude is amplified. The impact of noise injection in different layers can be seen in Fig.~\ref{fig:layer}.

\subsection{Stealth Analysis}

We analyzed the mean and standard deviation of the feature map across varying levels of noisy nodes and magnitude to assess the vulnerability of distributed CNNs in HC scenarios to adversarial attacks. For EdgeCNN on the MNIST dataset, with a number of impacted feature vector elements escalated to 50\%, the feature map's mean-variance and standard deviation-variance remained minimal, as shown in Tables~\ref{tab:stealth-analysis} and~\ref{tab:stealth-analysis2}. The noise magnitude was kept at 50\%. The LeNet-5 models on MNIST and Fashion datasets displayed an analogous pattern, with minimal fluctuations in their respective means and standard deviations, irrespective of the introduced noise magnitude. These results indicate that while major noise barely impacts feature statistics, it can drastically lower accuracy. This makes the attack stealthy while potentially evading the existing defense mechanisms, indicating the need for more robust countermeasures that consider such tactics.

\section{Problem Formulation}
\label{sec:problem_formulation}

In response to challenges observed in Section~\ref{sec:case_study} and the detected threats in Section~\ref{sec:threat_model}, we outline the architecture of the proposed framework and its deployment.
In HC, the CNN model is distributed across multiple edge nodes. Noise in any node can severely affect model performance. The objective is to develop a robust defense mechanism that maintains the system's output integrity across all data, layers, and nodes.

\textbf{Noise Detection:} The defense framework first examines node outputs to detect noise in intermediate feature maps, which is essential for the recovery process. A noise detection method \( \mathcal{D} \), compares the perturbed output \( M'(d, l_i, e_j) \) with the actual output \( M(d, l_i, e_j) \). If the difference exceeds a predefined threshold \( \theta \), the output is flagged as noisy. The threshold must be carefully set to balance false positives and false negatives in noise detection.

\textbf{Model Recovery:} Once the adversarial noise is detected, the next step is to recover the original feature maps to replace the adversarially perturbed ones bringing them as close as possible to the genuine output, hence ensuring the integrity of the system's overall performance. Let \( \mathcal{M}(d, l_i, e_j) \) represent the recovered output after applying the defense mechanism, which is a function of the perturbed output \( M'(d, l_i, e_j) \) and the detected discrepancy \( \Delta_{i,j} \).

\textbf{Objective Function:} The primary goal is to minimize the overall discrepancy between the recovered output and the genuine output across all data samples, layers, and edge nodes:
\begin{equation}
\min \sum_d \sum_{l_i \in L} \sum_{e_j \in E} \left| \mathcal{M}(d, l_i, e_j) - M(d, l_i, e_j) \right|
\label{eq:obj-func}
\end{equation}

\textbf{Subject to} the satisfaction of the following two constraints:
\begin{compactitem}
\item \textit{Detection Constraint:}
\begin{equation}
\begin{gathered}
\Delta_{i,j} = \mathcal{D}(M'(d, l_i, e_j), M(d, l_i, e_j)), \\
\forall d, \, l_i \in L, \, e_j \in E
\end{gathered}
\label{eq:detection_constraint}
\end{equation}
This ensures that the discrepancy between the actual and perturbed output is calculated for each data sample, layer, and edge node. It acts as a prerequisite for the recovery step, identifying the instances where the adversarial noise has affected the output (intermediate feature maps).

\item \textit{Recovery Constraint:}
\begin{equation}
\begin{aligned}
\forall d, &\, l_i \in L, \, e_j \in E, 
\mathcal{M}(d, l_i, e_j) = \\
& {\textrm{RecoverMethod}}(M'(d, l_i, e_j), \Delta_{i,j})
\end{aligned}
\label{eq:mitigation_constraint}
\end{equation}
%

Eq.~\ref{eq:mitigation_constraint} defines how to correct the detected noise. It specifies how the perturbed output is adjusted to minimize the discrepancy with the actual output, thus restoring the model’s accuracy compromised by the adversarial attacks. 

\end{compactitem}

\section{Proposed Framework for Secure Horizontal Edge with Adversarial Threat Handling} 
\label{sec:proposed_solution}

\begin{figure*}[!t]
    \centering
    \includegraphics[width=\textwidth]{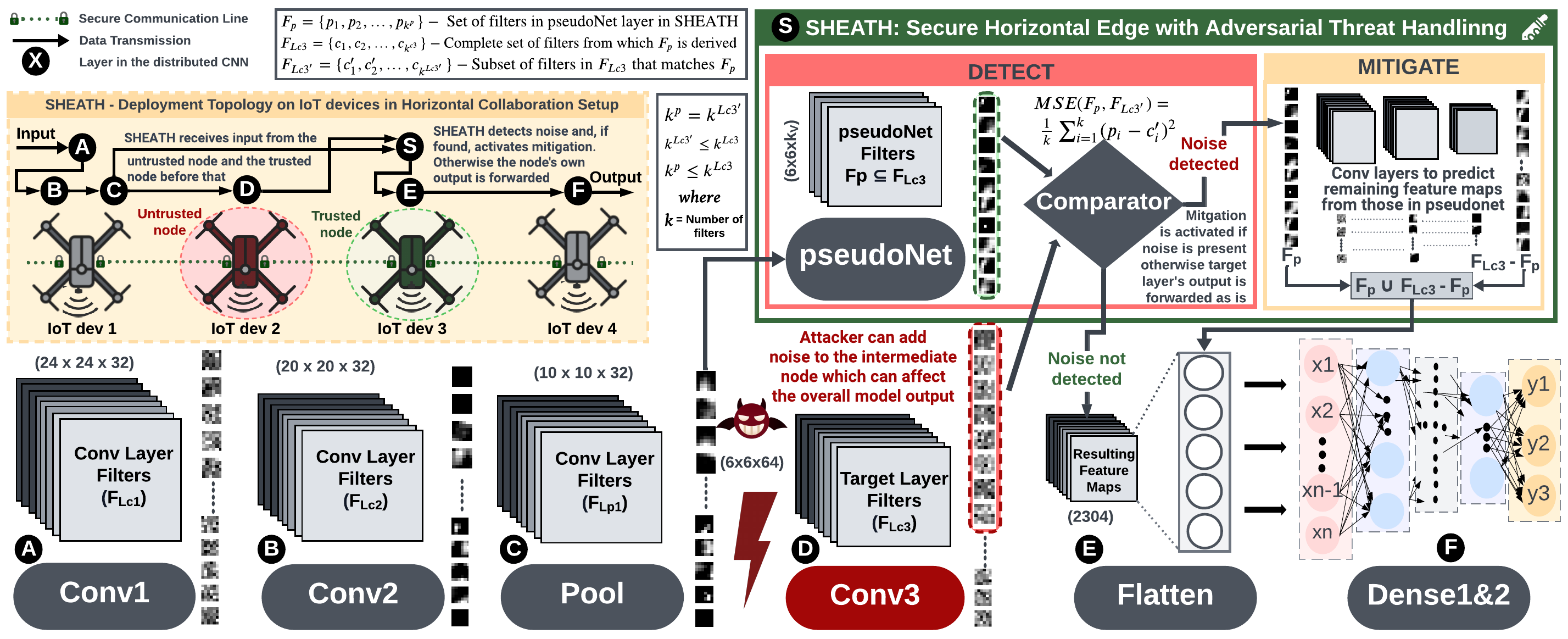}
    \caption{\small Overview of the SHEATH Framework. SHEATH has two modules: (i) \textit{Detect} and (ii) \textit{Recover}. It is deployed on a trusted node in HC-based edge devices to defend against adversarial noise from propagating to the rest of the CNN model. Integrated within drone-based trusted nodes, SHEATH secures the untrusted node (IoT dev2) by taking input from the preceding trusted node (IoT dev3). In the case of noise detection, SHEATH forwards the recovered output to the subsequent layer (IoT dev4). } 
    \vspace{-8pt}
   \label{fig:sheath}
\end{figure*}

We propose SHEATH, a novel framework to ensure the integrity and robustness of the CNN model deployed on HC edge devices with minimal overhead and latency, 
for resilience against adversarial attacks where not all the nodes are trusted in an HC architecture. 
The adversary can access the parameters of the model segment deployed on a compromised node, including the input and output feature maps. SHEATH addresses this by first detecting the presence of noise in the node's output. If the noise is detected, it corrects the output to the expected output, which it then provides to the next node.  SHEATH’s \textit{Detect} and \textit{Recover} modules are responsible for noise detection and feature map correction, respectively. The SHEATH’s \textit{Detect} consists of \textit{PseudoNet} and \textit{Comparator}. This is illustrated in Fig.~\ref{fig:sheath}, and the necessary technical details are also discussed in this section. By adding noise to these elements, the adversary can compromise the model's integrity.

\subsection{SHEATH-Detect}

This module is responsible for detecting the presence of noise in an untrusted node to ensure that it sends the correct output to the subsequent node. Given the node's untrusted nature, the \textit{Detect} module requires a reference set of noise-free feature maps generated by \textit{PseudoNet}. These feature maps are then sent to the \textit{Comparator} for verifying whether the difference between the noisy and non-noisy feature maps exceeds the predefined threshold.

\subsubsection{PseudoNet} 

The key idea for \textit{PseudoNet} is to have a secondary mechanism to recompute the expected feature maps from the untrusted node using fewer computational resources. The primary mechanism is the reference of comparison (non-noisy) in the \textit{Comparator}. Generation of reference feature maps can be achieved in two ways: (a) redundancy, which refers to the duplication of the whole targeted node, or (b) constructing a similar subset of the untrusted node with reduced hyperparameters (the \textit{PseudoNet} approach). The latter is a relatively lightweight mechanism. Let \( U \) be the untrusted node's layer(s), \( P \) the corresponding \textit{PseudoNet}'s layer(s), \( \theta_U \) the parameters in \( U \) (e.g., filters, neurons, kernel size), and \( \theta_P \) the parameters in \( P \). Then the formation of \textit{PseudoNet}'s layer(s) can be modeled as: 
\begin{equation}
\label{eq:pseudo}
\theta_P = \alpha \times \theta_U
\end{equation}

Where \( 0 < \alpha < 1 \) is the reduction factor, and \( \theta_P \subseteq \theta_U \), making \( P \) a computationally lighter counterpart of \( U \). This technique can be applied and tailored to \textit{PseudoNet} for any CNN layer \( L \). For example, if this reduction technique is applied to a convolutional layer, then \( L \) is denoted as \textit{conv}. The set of parameters, \( \theta_P \), consists of \( (\textit{W}_{\text{conv}}, \textit{b}_{\text{conv}}) \). Here, \( W_{\text{conv}} \) is a subset of filters from \( \theta_U \), and \( b_{\text{conv}} \) are the biases for those filters. The output is defined in Eq.~\ref{eq:output}.
\begin{equation}
\label{eq:output}
\text{Output}_P = f(W_{\text{conv}} \ast \text{Input} + b_{\text{conv}})    
\end{equation} 
where \( f \) is the activation function and \( \ast \) is the convolution. 

If the desired layer is the pooling Layer, \( L \) is represented as \textit{pool}. The set of parameters, \( \theta_P \)  (e.g., max, average), specify pooling operation and size. The output is defined in Eq.~\ref{eq:op}.
\begin{equation}
\label{eq:op}
\text{Output}_P = \text{pool}(\text{Input})
\end{equation}
If the target layer is the dense layer, \( L \) is represented as \textit{dense}. Set of parameters, \( \theta_P \), are \((\textit{W}_{dense}, \textit{b}_{dense})\). \( W_{\text{dense}} \) is a subset of weights from \( \theta_U \) and \( b_{\text{dense}} \) is a subset of biases from \( \theta_U \). In this case, the output will be
\begin{equation}
\text{Output}_P = f(W_{\text{dense}} \cdot \text{Input} + b_{\text{dense}})(\text{Input})
\end{equation}
where \( f \) is the activation function and \( \cdot \) is the dot product. 

For each layer type, the subset of parameters in \( \theta_P \) should reflect the lightweight aspect of \textit{PseudoNet} compared to \( \theta_U \). An appropriate \( \alpha \) can be chosen for each layer type based on the computational and security requirements. 

For brevity, we focus our explanation of the SHEATH framework on securing a convolutional layer within an untrusted node(s), as convolutional layers are fundamental to CNN architectures and crucial in defining the feature representation of input data. Their complexity, in terms of the number and size of filters and strides, makes them the primary target for attackers. We obtain this layer's \textit{PseudoNet} by reducing the number of filters/kernels. To illustrate, if  \( F_{Lc3} \) is the full set of filters in the original layer, then \( F_p \) represents the filters in our corresponding pseudoNet being a subset of \( F_{Lc3} \).

\noindent\textbf{\textit{Parameter reduction strategy in PseudoNet}}: 

Determining the optimal level of parameter reduction in models is a delicate process. A higher reduction of parameters can affect the accuracy, while a lower reduction can lead to computational inefficiencies. However, as a rule of thumb, the desired value of \(\alpha\), which lies between 0 and 1, can be empirically determined by iteratively testing the reduced model's performance and comparing it to the baseline, aiming to achieve a balance between accuracy and efficiency.

\subsubsection{Comparator} 

After the \textit{PseudoNet} computes the reference feature maps for the \textit{Comparator}, the next step is to compare these maps with the noisy feature maps produced by the target layer in the untrusted node. The primary method of this comparison is computing the mean squared error (MSE)  between the noisy and non-noisy feature maps. As \textit{PseudoNet} is a reduced subset of the untrusted node, only the corresponding feature maps from both are compared, which makes noise detection computationally efficient. Given two sets of filters, one from the \textit{PseudoNet} represented as \( F_p \) and the other, a subset from the target layer represented as \( F_{Lc3'} \) (which corresponds to \( F_p \)), the MSE is defined as:
\begin{equation}
\label{eq:mse-comp}
MSE(F_p, F_{Lc3'}) = \frac{1}{k} \sum_{i=1}^{k} (p_i - c'_i)^2
\end{equation}
where \( p_i \) represents the filter values from the \textit{PseudoNet}, \( c'_i \) represents the corresponding filter values from the subset of the target layer, \( F_{Lc3'} \), and \( k \) is the total number of filters being compared. If the computed MSE surpasses a predefined threshold, it indicates that there is noise in the system. 
Subsequently, measures to correct this noisy output are activated to ensure the overall system integrity.

\noindent\textbf{\textit{Threshold Selection Strategy}}: The threshold for MSE is selected based on the comparison performed during the sanity check conducted in a noise-free environment. Here, the feature maps from the \textit{PseudoNet} are compared against those from the target layer of the base model. In an ideal, noise-free situation, the MSE between the \textit{PseudoNet} and the base model's target layer should be 0 or very close to it. However, in real-world scenarios, some variation is expected. For our case, the sanity check yielded an MSE of 0, which we used as the threshold.

\subsection{SHEATH-Recover}

The primary role of this module is to ensure that accurate and noise-free output is delivered to the subsequent nodes in the system. If the \textit{Detect} module determines that there is no noise in the feature maps, the output of the untrusted node is forwarded as-is to the subsequent node. However, if noise is detected, the \textit{Recover} module detects and corrects noise, forwarding the output to the next node. It trains convolutional layers by using feature maps from \textit{PseudoNet}. Feature maps are merged based on filter index, e.g., if the untrusted node generates 64 feature maps and \textit{PseudoNet} yields five, \textit{Recover} predicts the remaining 59 feature maps. The merged output, consisting of the predicted and \textit{PseudoNet} feature maps, is forwarded to the subsequent node, ensuring a noise-free output. Mathematically, \( F_{LC3} \) is the complete set of filters in the target layer of the untrusted node, \( F_{p} \) is a subset of filters in \( F_{LC3} \) that matches \textit{PseudoNet}, \( n \) is the total number of feature maps in the target layer of the untrusted node, and \( p \) is the number of feature maps produced by \textit{PseudoNet}, where \( p \leq n \). The objective, then, is to generate the missing \( (n - p) \) feature maps using the convolutional layers in the \textit{Recover} module and merge them with the \( p \) from \textit{PseudoNet}.

\noindent\textbf{Generating Missing Feature Maps}: This can be done as:
\begin{equation}
\label{eq:gen-fms}
FM_{missing} = Conv_{l,f}(Input_{Recover})
\end{equation}
Where \(FM_{missing}\) represents the set of missing feature maps, \(F_{LC3} - F_{p}\). \(Conv_{l,f}\) is the convolution operation using the optimal \(l\) and \(f\) obtained from the grid search, and \(Input_{Recover}\) is the input to the \textit{Recover} module. 

\noindent\textbf{Merging Feature Maps}: The merged feature maps, taking the union of those from the \textit{PseudoNet} and \textit{Recover} modules are:
\begin{equation}
\label{eq:merged-fms}
FM_{merged} = F_{p} \cup (F_{LC3} - F_{p})
\end{equation}
Thus, the final \(FM_{merged}\) is a union of the feature maps from \textit{PseudoNet} and the missing ones produced by the \textit{Recover} module to match the target \(F_{LC3}\). The training for these convolutional layers is executed offline, ensuring no real-time delays or performance hitches. The architecture of the \textit{Recover} module, including the number of layers and filters within these layers, is determined optimally through a grid search algorithm. The primary objective is a tradeoff between accuracy and overhead. The grid search function is given by
%
\begin{equation}
\label{eq:grid-search}
G(\mathit{grid}) = \min_{1 \leq l \leq n, 1 \leq f \leq m}~\text{Loss}(\mathit{Model}_{l,f})
\end{equation}
Where \(G\) is the grid search function over the defined \(grid\), comprising all possible combinations of the number of convolutional layers \(l\) and number of filters \(f\) in each layer. Eq.~\ref{eq:grid-search} represents the grid search function where 
\(l\) varies from 1 to 
\(n\) (number of layers) and \(f\) varies from 1 to 
\(m\) (number of filters per layer), and the goal is to minimize the loss function for the given model configuration. \text{Loss}(\(Model_{l,f}\)) represents the loss function of the model configuration trying to match \(F_{LC3} - F_{p}\) as accurately as possible.

\subsection{Deployment Limitations}

Consider a sequence of edge nodes \(E = \{e_1, e_2, ... e_x\}\). For the sake of brevity, we consider each node has one layer of the CNN model to discuss the feasibility of SHEATH deployment in three scenarios: (i) single-layer attack, (ii) non-consecutive multi-layer attacks, and (iii) consecutive multi-layer attacks.

\begin{figure}[!t]
    \centering
    \includegraphics[width=0.8\columnwidth]{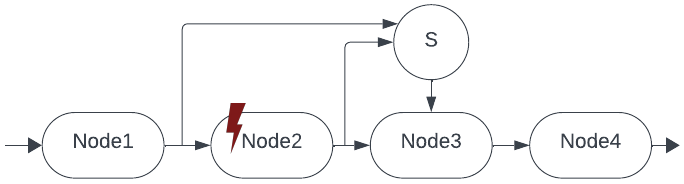}
    \caption{Single-layer attack in HC-based edge devices.}
    \label{fig:dep-1}
\end{figure}

\begin{figure}[!t]
    \centering
    \includegraphics[width=0.8\columnwidth]{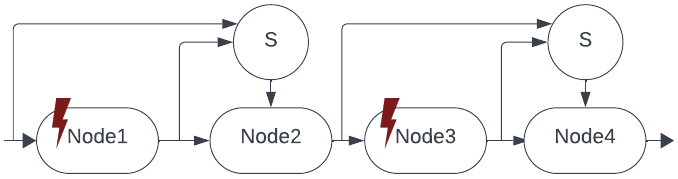}
    \caption{Non-consecutive multi-layer attack in HC edge devices.}
    \label{fig:dep-2}
\end{figure}
\textbf{Single Layer Attack:}
If \(e_i\) is untrusted, SHEATH is deployed on \(e_{i+1}\) to verify the computations performed by \(e_i\) as shown in Fig.~\ref{fig:dep-1}. For instance, in a setup of nodes \(E = \{e_1, e_2, e_3, e_4, e_5\}\), if \(e_2\) becomes compromised, then SHEATH is on \(e_3\), ensuring \(e_3\) validates and possibly rectifies computations from \(e_2\) before progressing with its tasks.

\textbf{Non-consecutive Multi-layer Attacks:}
For several non-consecutive compromised nodes, say \(e_i\) and \(e_j\) (where \(|i-j| > 1\)), SHEATH is deployed on \(e_{i+1}\) and \(e_{j+1}\) respectively as can be seen in Fig.~\ref{fig:dep-2}. Reflecting on a system with nodes \(E = \{e_1, e_2, e_3, e_4, e_5, e_6\}\), if nodes \(e_2\) and \(e_4\) are both untrusted, SHEATH is deployed on \(e_3\) and \(e_5\). These nodes will then be responsible for individually confirming and potentially amending the computations of \(e_2\) and \(e_4\).

\begin{figure}[t]
    \centering
    \includegraphics[width=0.8\columnwidth]{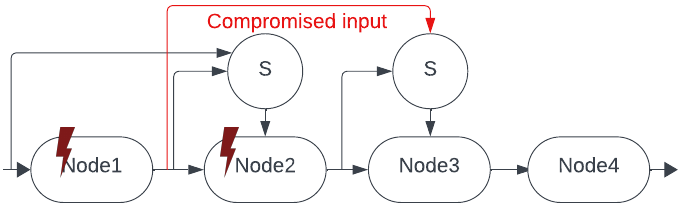}
    \caption{Consecutive multi-layer attack in HC edge devices.}
    \vspace{-9pt}
    \label{fig:dep-3}
\end{figure}

\textbf{Consecutive Multi-layer Attacks:}
As seen in Fig.~\ref{fig:dep-3}, when two consecutive nodes, \(e_i\) and \(e_{i+1}\), are under threat, SHEATH applied on \(e_{i+2}\) may receive compromised input, reducing its efficacy. Taking \(E = \{e_1, e_2, e_3, e_4\}\), if \(e_2\) and \(e_3\) are both untrusted, SHEATH on \(e_4\) may be ineffective, as it gets potentially flawed data from \(e_3\).

\textbf{Limitations:} The deployment of SHEATH provides a mechanism to secure the system, but it isn't without challenges. One significant limitation is the susceptibility to consecutive multi-layer attacks, as seen in Fig.~\ref{fig:dep-3}. In these cases, even if SHEATH is in place, it may operate on flawed data. This results in a cascading effect of the noise compromising the rest of the model. For SHEATH to be effective, it operates assuming it receives trustworthy input. In situations where \( e_i \) is compromised, SHEATH on \( e_{i+1} \) is designed to handle and rectify the potential noise or alterations. However, when both \( e_i \) and \( e_{i+1} \) are compromised, the SHEATH on \( e_{i+2} \) faces a dilemma. The data it receives has passed through two layers of potential malicious alterations. This hampers its ability to distinguish actual data from the compromised one. Thus, in such cases, we can reconsider the distribution of nodes to avoid model deployment on two consecutive untrusted nodes.

\subsection{Theoretical Complexity Analysis for SHEATH}
\label{subsec:complexity}
The other alternative for security in HC-based CNN architectures would be to do a redundancy, i.e., duplicating the targeted untrusted node(s) (one node redundancy for single-layer and multiple node redundancy under a multi-layer attack(s)). The theoretical computational complexity for SHEATH and doing a complete redundancy can be represented as follows. Let \(t_d\), \(t_m\), and \(t_r\) be the times and \(m_{\text{d}}\), \(m_{\text{m}}\), and \(m_{\text{r}}\) be the memories for SHEATH-\textit{Detect}, SHEATH-\textit{Recover}, and redundancy, respectively. The complexity for SHEATH is typically \( O(t_d) \), occasionally peaking at \( O(t_d + t_m) \) upon noise detection. Contrarily, redundancy consistently operates at \( O(t_r) \). If \( t_d \) is minimal, \( t_m \) is infrequent, and \( t_r \) has significant overhead, then the peak complexity of SHEATH is likely less than redundancy as in Eq.~\ref{eq:complexity}.
\begin{equation}
\label{eq:complexity}
O(t_d + t_m) < O(t_r)
\end{equation}
Considering resource efficiency, SHEATH primarily utilizes resources at \( O(t_d) \), with occasional rises, whereas redundancy demands \( O(t_r) \) continuously. This implies higher long-term costs for redundancy. In latency, SHEATH incurs \( O(t_d) \), only rising upon noise detection, while redundancy is consistently \( O(t_r) \), potentially higher due to synchronization or switching. 

For overhead, SHEATH mainly correlates with \( O(t_d) \), with occasional costs of \( O(t_m) \). Redundancy, however, consistently incurs \( O(t_r) \) due to backup system maintenance. In conclusion, SHEATH offers a more efficient paradigm than redundancy, especially with infrequent noise detections. We experimentally validate the complexity for both SHEATH and redundancy in RQ8 of Section~\ref{sec:evaluation}.

\section{Performance Evaluation}
\label{sec:evaluation}

To evaluate the performance of SHEATH across diverse CNN configurations, we experiment with three CNN models: \textit{LeNet}, \textit{MiniVGGNet}, and \textit{EdgeCNN}, our custom test-case CNN. We evaluate both SHEATH-\textit{Detect} and SHEATH-\textit{Recover} to determine whether noise is being accurately detected and the feature maps are being corrected. We use several metrics for this: (1) \textit{Accuracy}, which measures its ability to identify feature maps with adversarial noise correctly; (2) \textit{Precision}, representing the proportion of correctly identified noisy feature maps out of all classified as noisy; (3) \textit{Recall}, indicating the success in identifying actual noisy feature maps out of all present; and (4) \textit{F1-score}, which balances noise detection and reduces false positives as the harmonic mean of recall and precision. 
We investigated and conducted experiments for the following research questions (RQs).


\textbf{RQ1}: How effectively does SHEATH detect adversarial noise across different layers of varied CNN architectures?

\textbf{RQ2}: What is the performance across different datasets?

\textbf{RQ3}: How does the number of filters and other hyperparameters influence SHEATH's detection performance?

\textbf{RQ4}: What are the false positives and false negatives?

\textbf{RQ5}: How does the noise detection accuracy of SHEATH vary with the strength of adversarial noise?


\textbf{RQ6}: What is the performance with model complexity?

\textbf{RQ7}: Does the location of noise injection in the model impact SHEATH's performance?

\textbf{RQ8}: What computational overhead is added by SHEATH-\textit{Detect} and SHEATH-\textit{Recover}, and how does the complexity compare with that of a complete redundancy?

\textbf{RQ9}: What is SHEATH's efficacy when multiple non-consecutive nodes are attacked?

\textbf{RQ10}: What is SHEATH's efficiency in recovering original feature maps with different levels of noise injection?

\subsection{Evaluation Results}

This subsection entails the answers to the aforementioned RQs with experimental results.



            

\begin{figure}[!t]
\centering
        \subfigure[\label{fig:rq1}]
        {
        \includegraphics[width=0.47\columnwidth, height=3.7cm]{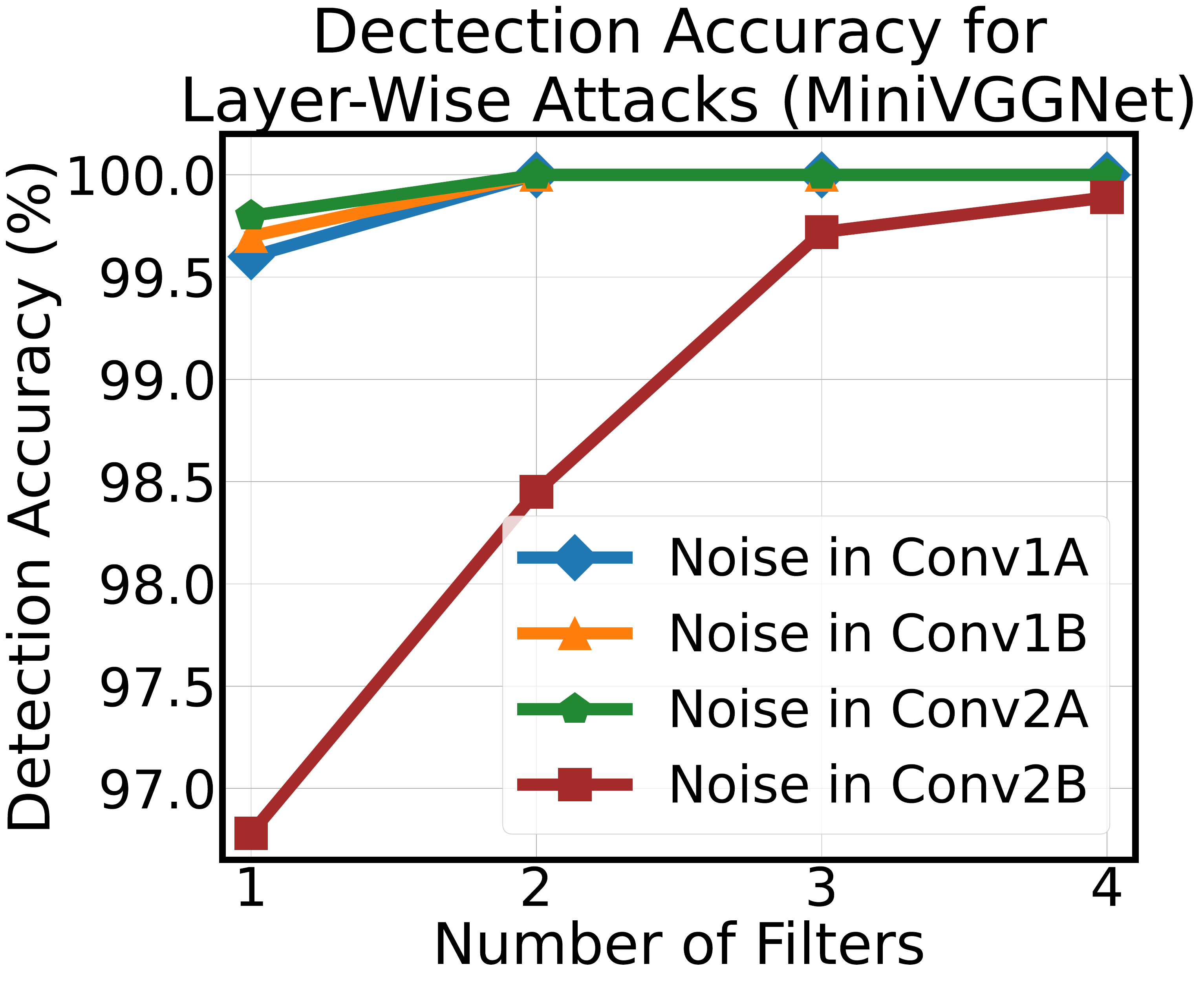}
        }\hspace{-6pt}
        \subfigure[\label{fig:rq1-b}
            ]
        {
        \includegraphics[width=0.47\columnwidth]{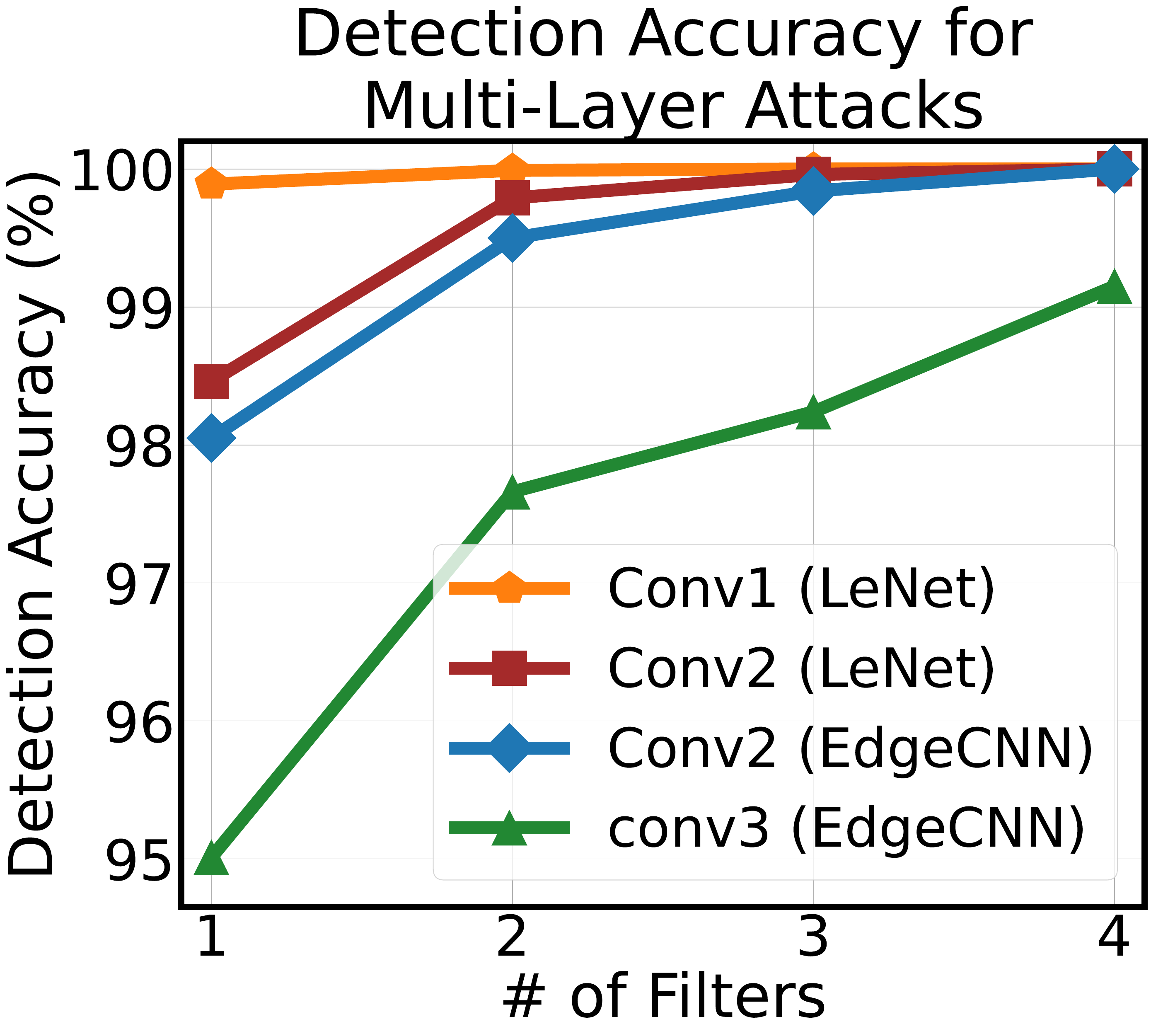}
        }
\vspace{-10pt}
\caption{\small SHEATH-detect's Efficacy across Different Layers in CNN Architectures (a) MiniVGGNet and (b) EdgeCNN.}
\label{fig:rq5}
\vspace{-6pt}
\end{figure}

\vspace{3pt}
\noindent\textbf{\textit{RQ1 - Analysis of SHEATH's Detection Efficacy across Different Layers in CNN Architectures:}} Analyzing CNN layers' resilience to noise reveals vulnerabilities. Deeper LeNet layers suffered more accuracy drop, while MiniVGGNet's \textit{``Conv2A"} had a significant dip. SHEATH's evaluation on MiniVGGNet indicated high accuracy in early layers. Notably, \textit{``Conv2B"} accuracy jumped from 89.83\% with one filter to 99.89\% with five, emphasizing filter analysis in noise detection. Simultaneous attacks on multiple layers of \textit{EdgeCNN} in \textit{``Conv2"} and \textit{``Conv3"}, with each layer having its detection mechanism, showed considerable detection performance with accuracy increasing from 95\% with one filter to 99.01\% with five for \textit{``Conv3"}.
SHEATH's consistent high detection, represented in Fig.~\ref{fig:rq1} and \ref{fig:rq1-b}, underscores its defense capability against noise, regardless of the injection point.

        

\begin{figure}[t] 
    \centering    
    \includegraphics[width=0.75\columnwidth]{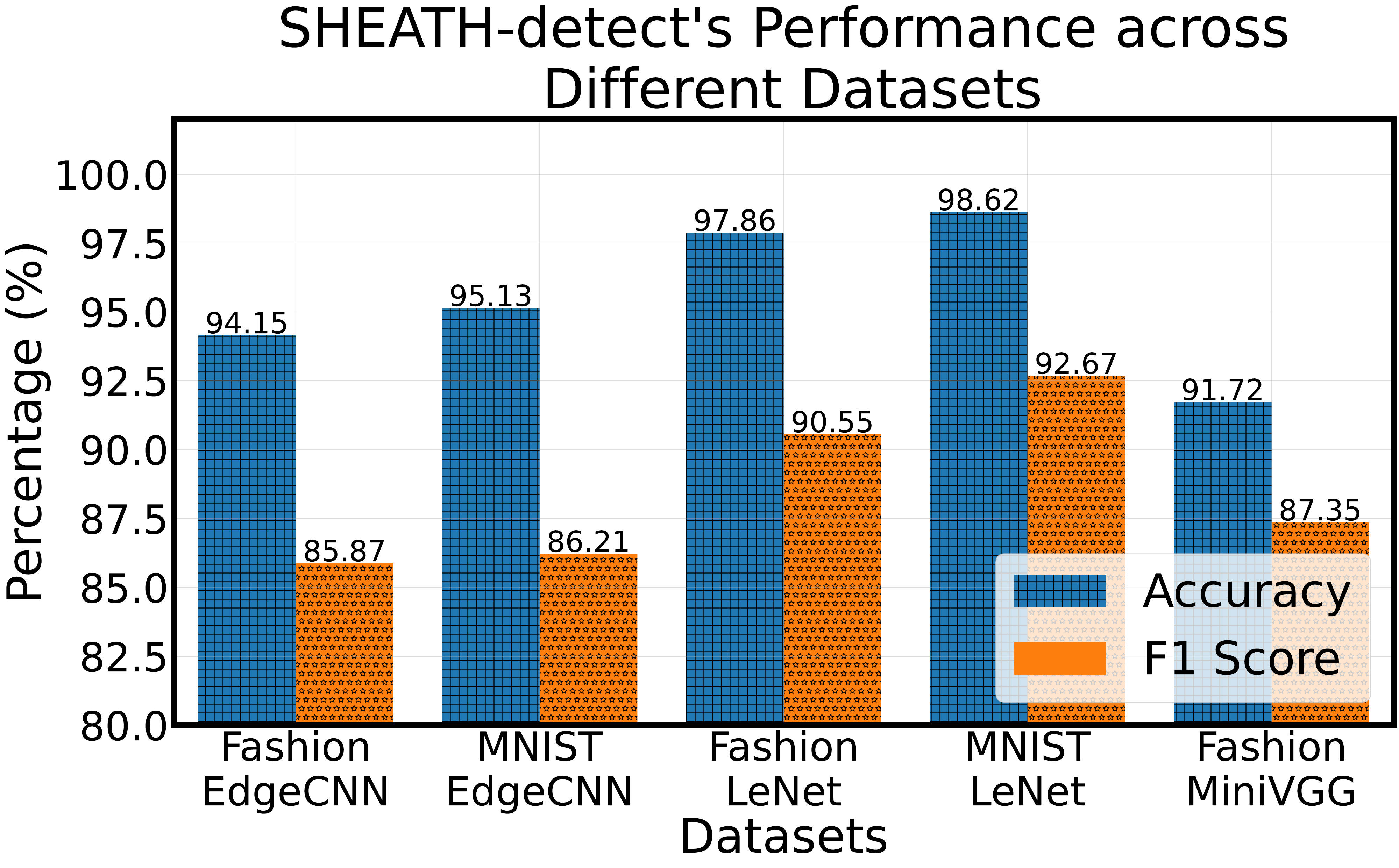}
    \vspace{-6pt}
    \caption{\small SHEATH's Performance with varied datasets.}
    \label{fig:rq7}
    \vspace{-6pt}
\end{figure}

\vspace{3pt}
\noindent\textbf{{\textit{RQ2 -SHEATH's Performance with Varied Datasets:}}} Assessing the adaptability of an adversarial noise detection mechanism across varied datasets is vital for understanding its robustness and real-world relevance. We evaluated SHEATH across representative datasets using a single filter. For MiniVGGNet on Fashion, accuracy and F1 score stood at 91.72\% and 87.35\%. For LeNet on the Fashion dataset, the values were 97.86\% and 90.55\%, while on MNIST, they reached 98.62\% and 92.67\%. As illustrated in Fig.~\ref{fig:rq7}, SHEATH performs well across datasets. The variance in performance suggests that dataset characteristics can influence SHEATH's efficacy. While SHEATH showcases broad adaptability, customizing it for specific datasets might further optimize its capabilities.
\vspace{3pt}

\begin{figure}[!t]
\centering
        \subfigure[\label{fig:rq3-1}]
        {
        \includegraphics[width=0.47\columnwidth]{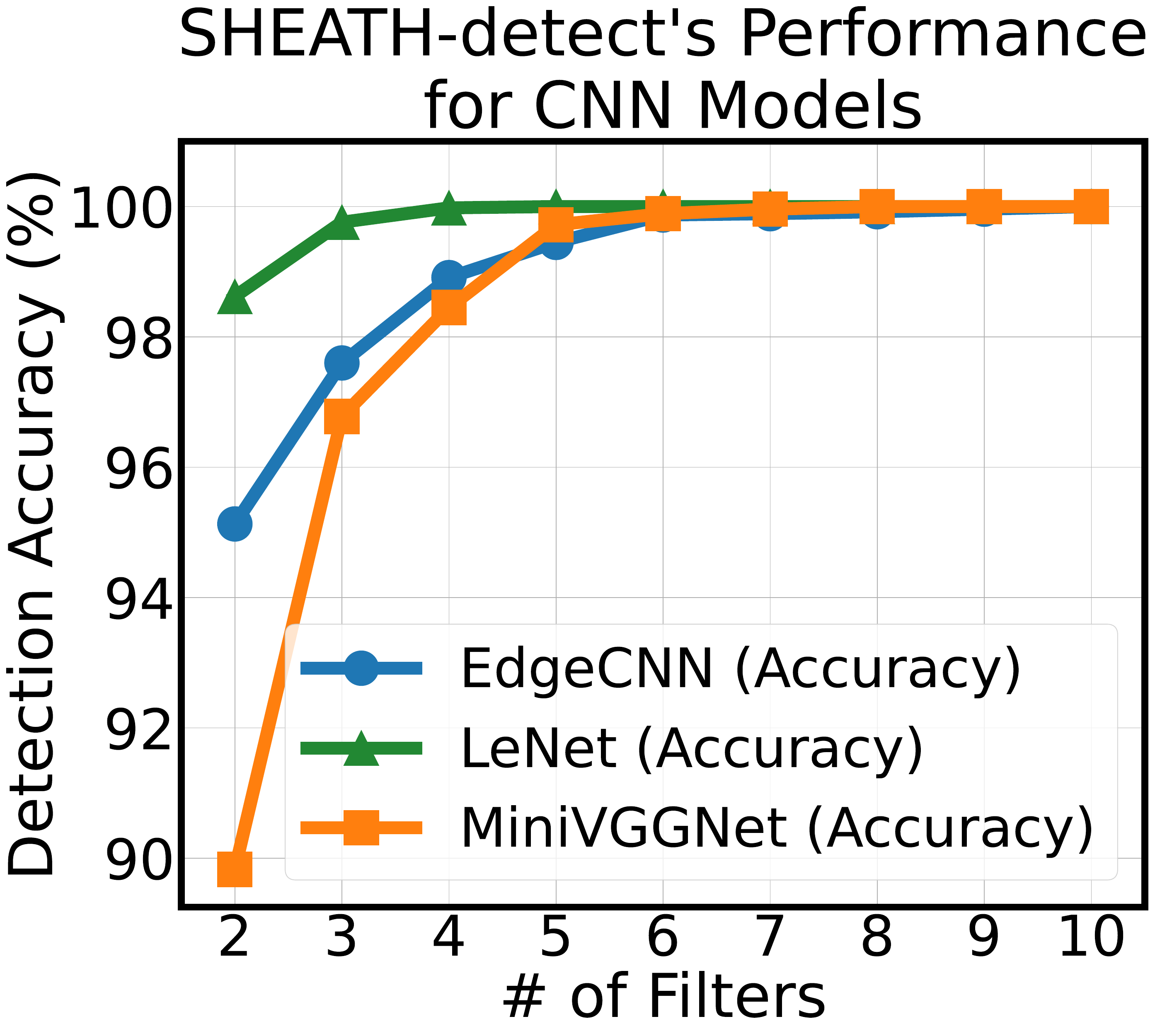}
        }\hspace{-6pt}
        \subfigure[\label{fig:rq3-2}
            ]
        {
        \includegraphics[width=0.47\columnwidth]{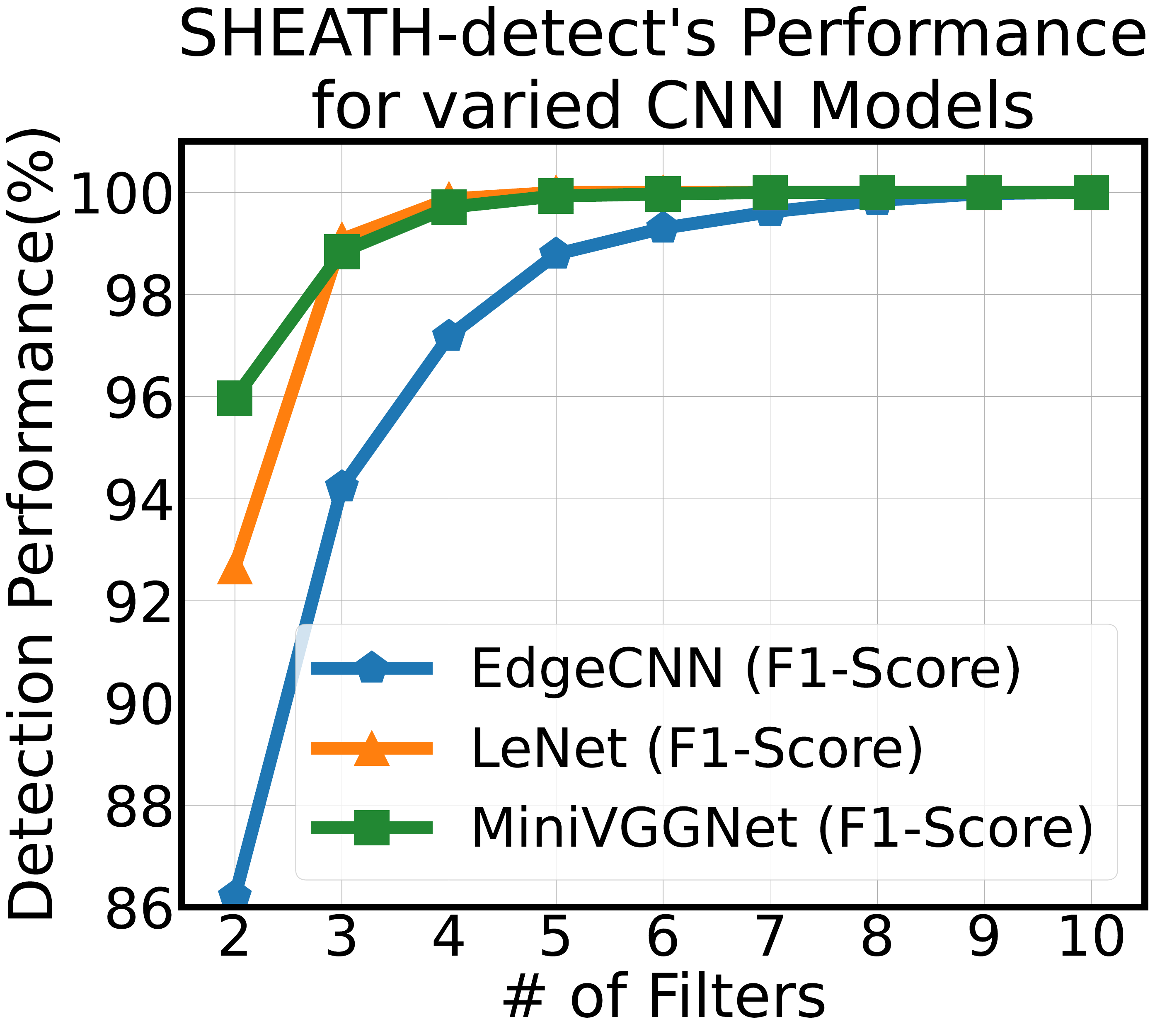}
        }
\vspace{-6pt}
\caption{\small Influence of Hyperparameters on SHEATH: (a) Accuracy and (b) F1-Score when no. of filters are varied in \textit{PseudoNet}}.
\label{fig:rq3}
\vspace{-15pt}
\end{figure}

\begin{table}[!t]
    \centering
    \caption{False positives and false negatives in VeriNET.}
    \label{tab:rq4}
        \begin{tabular}{|c|c|c|c|c|}
    \hline
        CNN Model & Dataset & False Positive & False Negative & ROC \\ \hline
        EdgeCNN & MNIST & 0 & 150 & 0.925 \\ \hline
        LeNet & MNIST & 0 & 42 & 0.979 \\ \hline
        MiniVGGNet & CIFAR-10 & 0 & 370 & 0.962 \\ \hline
    \end{tabular}
    \vspace{-9pt}
\end{table}

\vspace{3pt}
\noindent\textbf{{\textit{RQ3 - Influence of Hyperparameter Variations on SHEATH-\textit{Detect}'s Performance across CNN Architectures:}}} In deploying CNN-based systems like SHEATH, hyperparameter choices like filter count, significantly impact performance. We hypothesized that SHEATH can detect adversarial noise even with a single affected filter, but balancing filter count for improved detection while considering computational overhead is crucial. Experimental results showed \textit{EdgeCNN}'s detection accuracy starting at 95.13\% for one filter, reaching 100\% by ten filters, with a similar trend in the F1-score. For LeNet, accuracy began at 98.62\%, hitting 100\% by three filters, maintaining this trend. MiniVGGNet started at 89.83\%, achieving 100\% accuracy by seven filters. These findings emphasize SHEATH's effectiveness with few filters and the importance of optimizing for computational efficiency, as seen in Fig.~\ref{fig:rq3}.


\vspace{3pt}
\noindent\textbf{\textit{RQ4 - Evaluation of SHEATH's performance in false positives and false negatives across CNN models and datasets:}} Evaluating SHEATH's adversarial noise detection also involves assessing false positives and negatives. Table.~\ref{tab:rq4} outlines these metrics for SHEATH across varied CNN models and datasets. Specifically, on the MNIST dataset, \textit{EdgeCNN} showed zero false positives but had 150 false negatives. LeNet, on the same dataset, recorded zero false positives and only 42 false negatives. For the CIFAR-10 dataset, MiniVGGNet had zero false positives but a larger 370 false negatives. The findings underscore the necessity to refine SHEATH's detection, reducing false negatives for consistent accuracy across varied CNNs and datasets. It's vital to optimize SHEATH's capabilities for enhanced noise detection across diverse CNNs.


\begin{figure}[!t]
\centering
        \subfigure[\label{fig:rqfive1}]
        {
        \includegraphics[width=0.47\columnwidth]{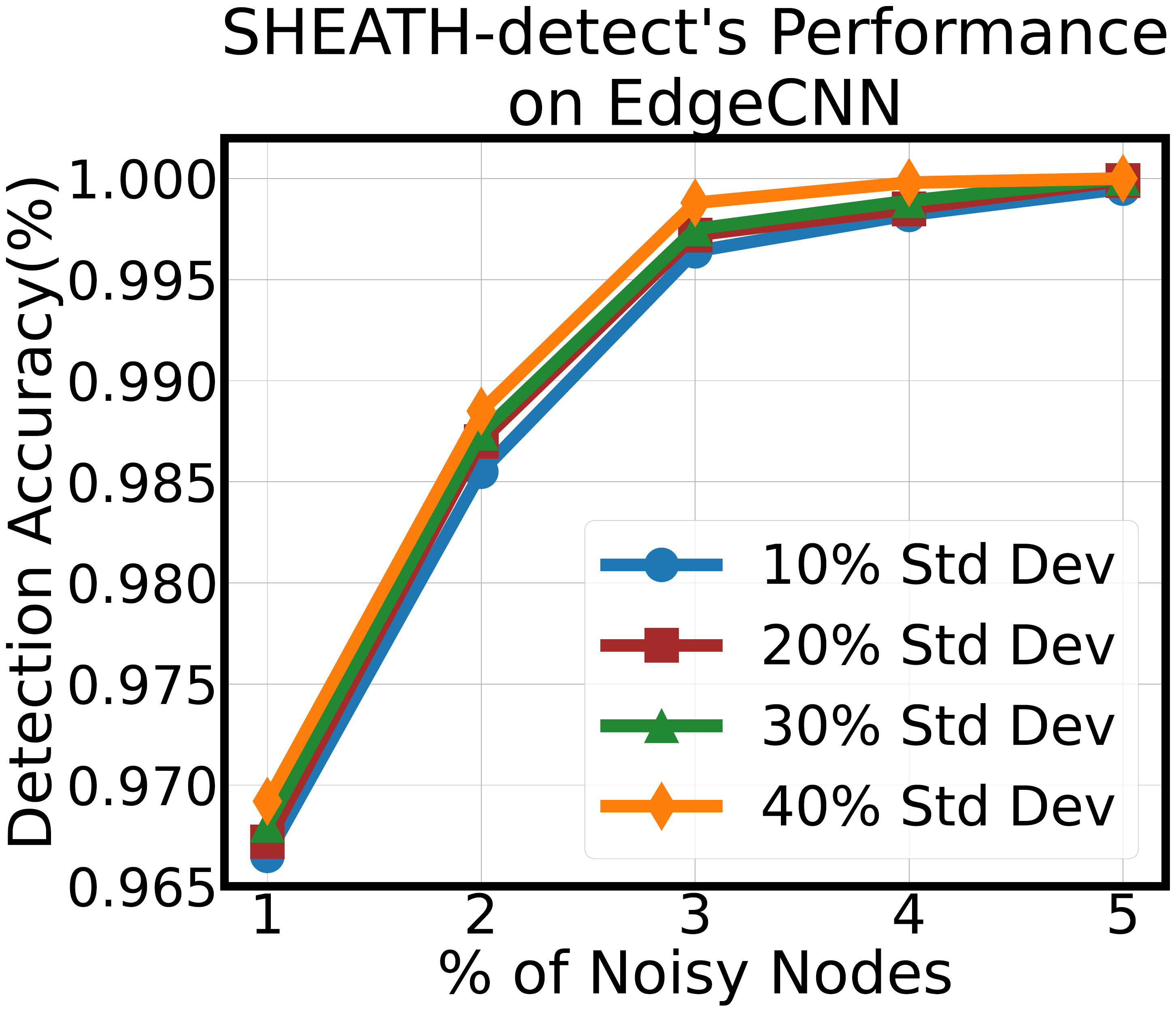}
        }\hspace{-6pt}
        \subfigure[\label{fig:rqfive2}
            ]
        {
        \includegraphics[width=0.47\columnwidth]{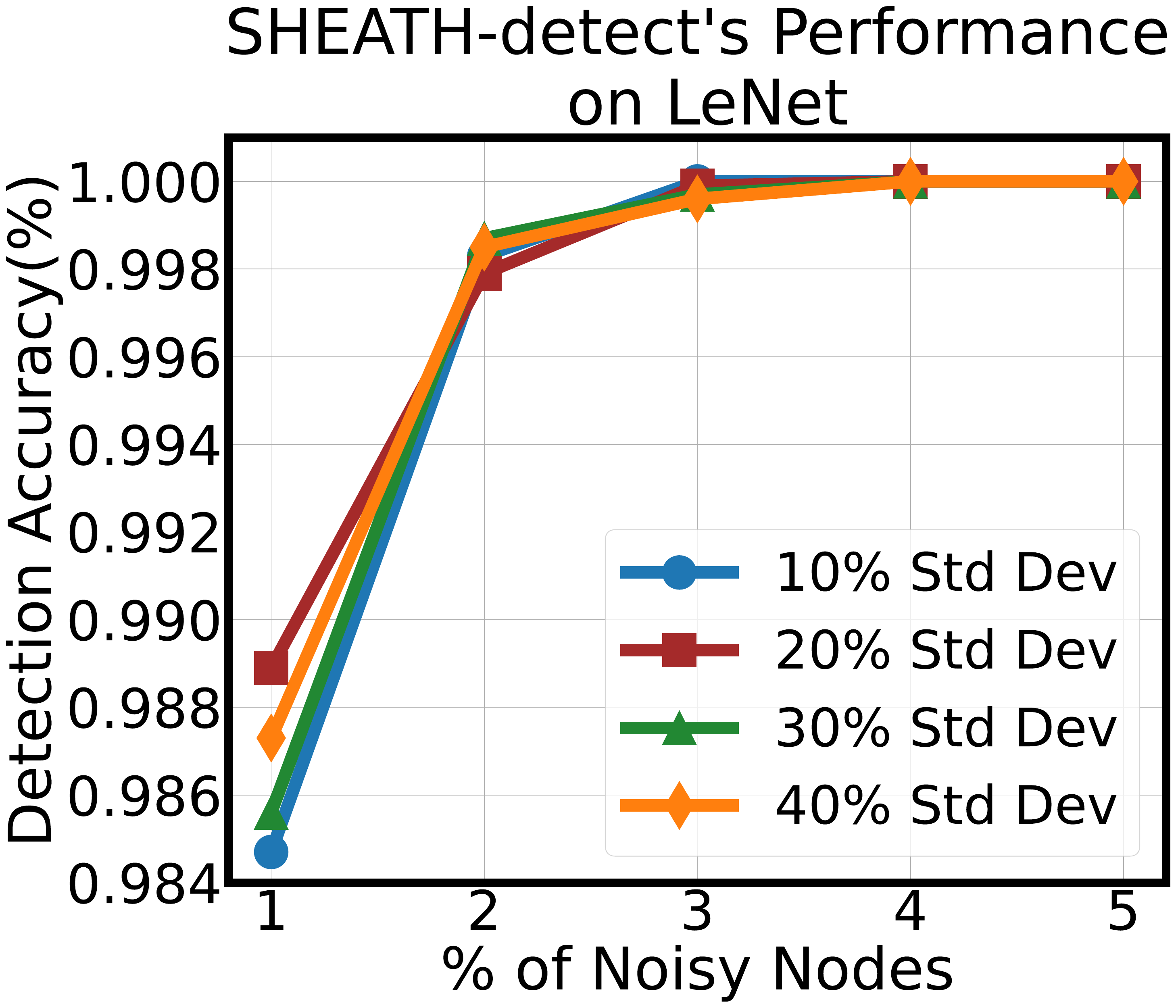}
        }
        \subfigure[\label{fig:rqfive3}
            ]
        {
        \includegraphics[width=0.47\columnwidth]{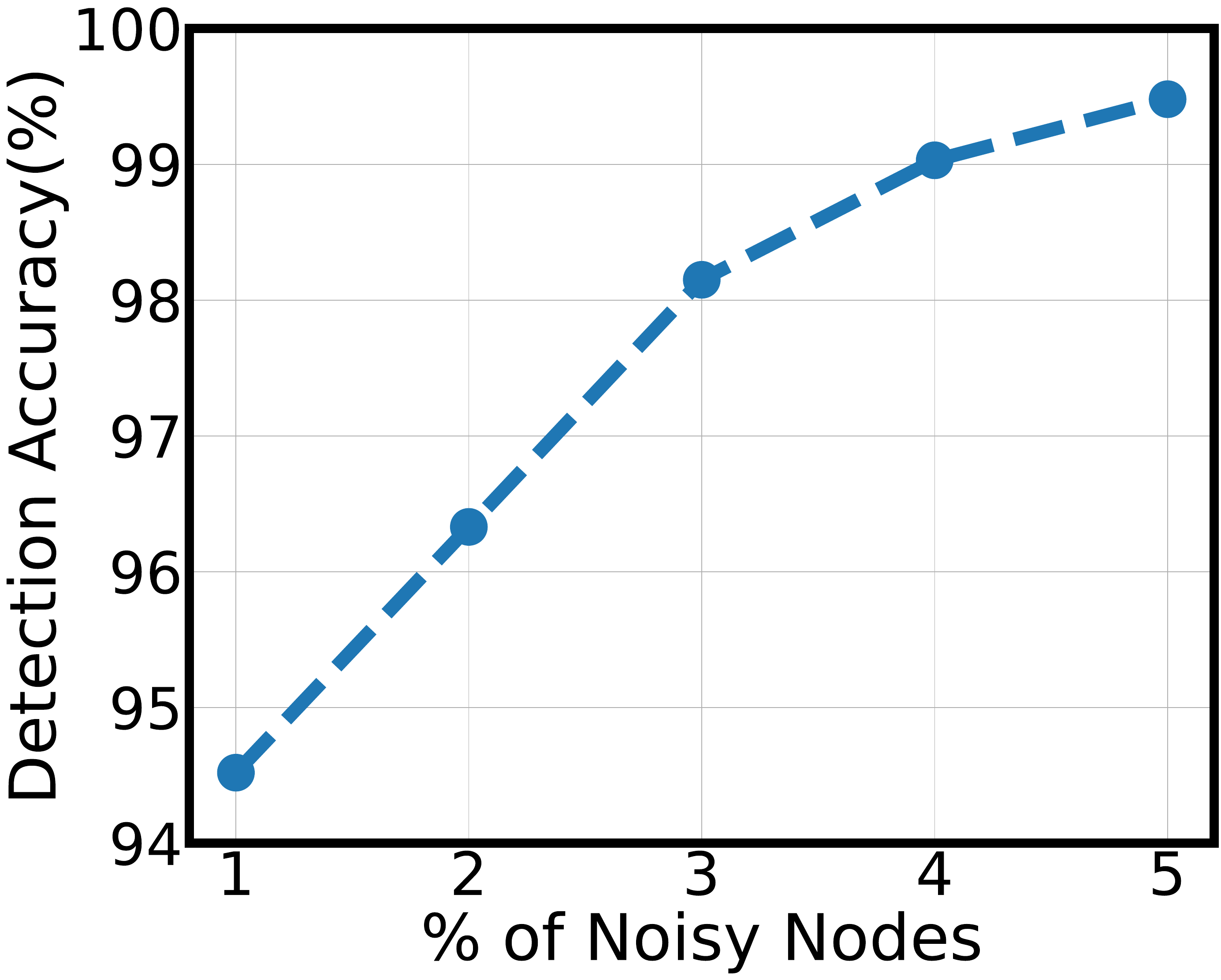}
        }\hspace{-5pt}
        \subfigure[\label{fig:rqfive4}
            ]
        {
        \includegraphics[width=0.47\columnwidth]{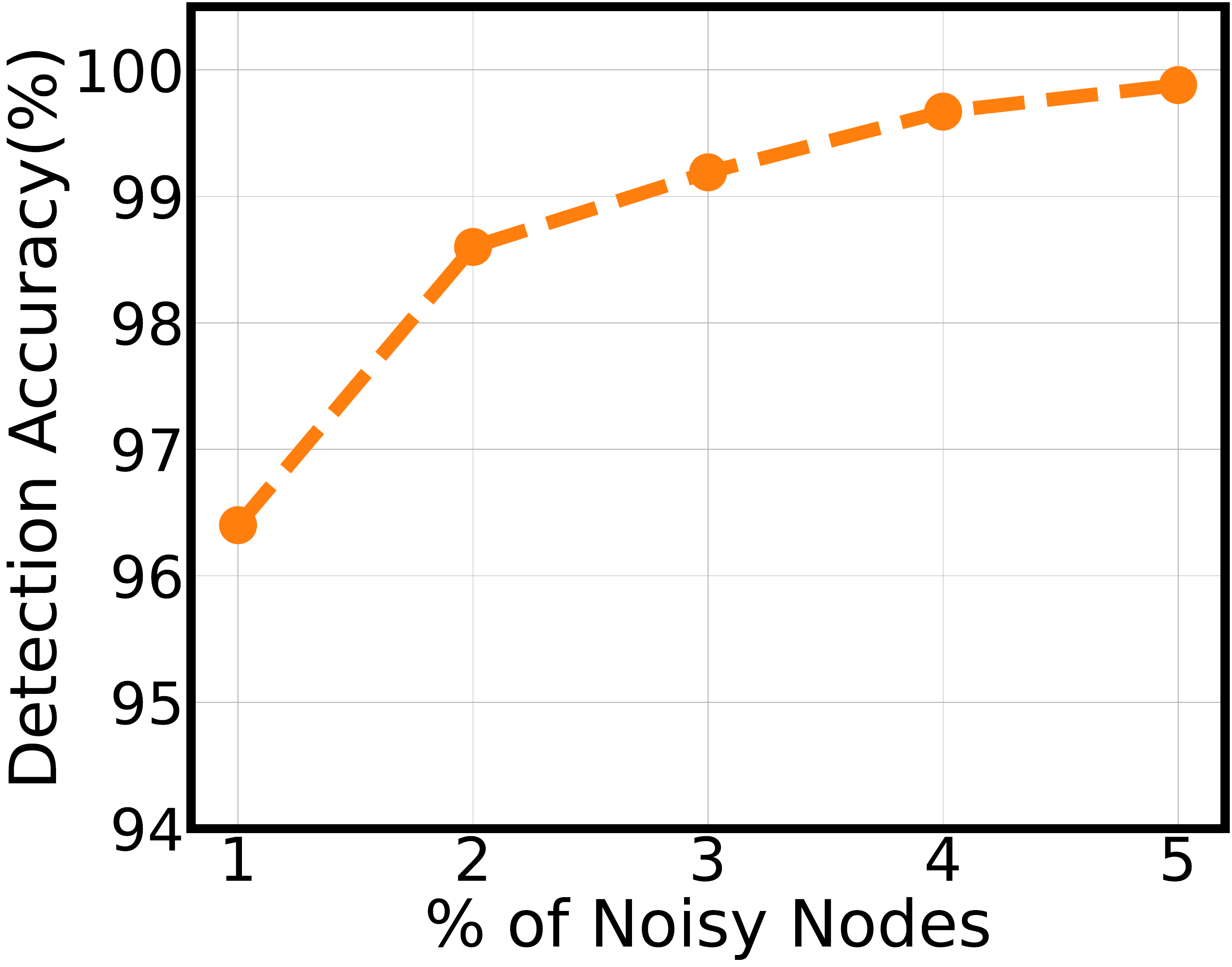}
        }
\vspace{-6pt}
\caption{\small SHEATH-\textit{Detect}'s performance against varied adversarial strengths in (a) EdgeCNN and (b) LeNet for Gaussian Noise and (c) EdgeCNN and (d) LeNet for Polarity Switch Attack}
\vspace{-12pt}
\end{figure}

\vspace{3pt}
\noindent\textbf{{\textit{RQ5 - Robustness Against Varied Noise Characteristics:}}} To assess SHEATH's detection capabilities against different adversarial noise attacks, we evaluated it under Gaussian noise and polarity switch attacks. The polarity switch attack involves inverting the sign values within the feature vector of the CNN model, changing positive values to negative and vice versa. This manipulation affects only the signs, not the magnitudes of the values. The severity of this perturbation is measured by the percentage of noisy nodes, indicating the portion of the feature vector impacted by this adversarial technique. Such attacks typically have a more pronounced effect on the CNN architecture's memory-centric or fully connected layers.
For \textit{EdgeCNN}, accuracy also improved as Gaussian noise levels augmented. Specifically, with a noise level of 1\% and a standard deviation of 10\%, the accuracy stood at 0.9660. However, with a 5\% noise increase, the accuracy impressively surged to 0.9998. This pattern was consistent across all standard deviation values.
Additionally, when subjected to the polarity switching, \textit{EdgeCNN} exhibited robustness. At a noise level of 1\%, the detection accuracy was 94.52\%, showcasing the model's resilience to this specific adversarial perturbation. Moreover, at a noise level of 5\%, the detection accuracy further increased to 99.48\%.
In the evaluation of LeNet, the model initially achieved an accuracy of 98.47\% with 1\% noise and a 10\% standard deviation. The accuracy improved to 100\% with Gaussian noise levels of 3\% and 4\%. In tests involving polarity switching, the model's accuracy was 96.40\% at a noise level of 1\%, and increased to 99.88\% at 5\% noise. These results demonstrate that the SHEATH framework effectively maintains high detection accuracy across varying levels of adversarial noise, as illustrated in Fig.~\ref{fig:rqfive1} -~\ref{fig:rqfive4}.

\begin{figure}[t] 
    \centering
    \includegraphics[width=0.75\columnwidth]{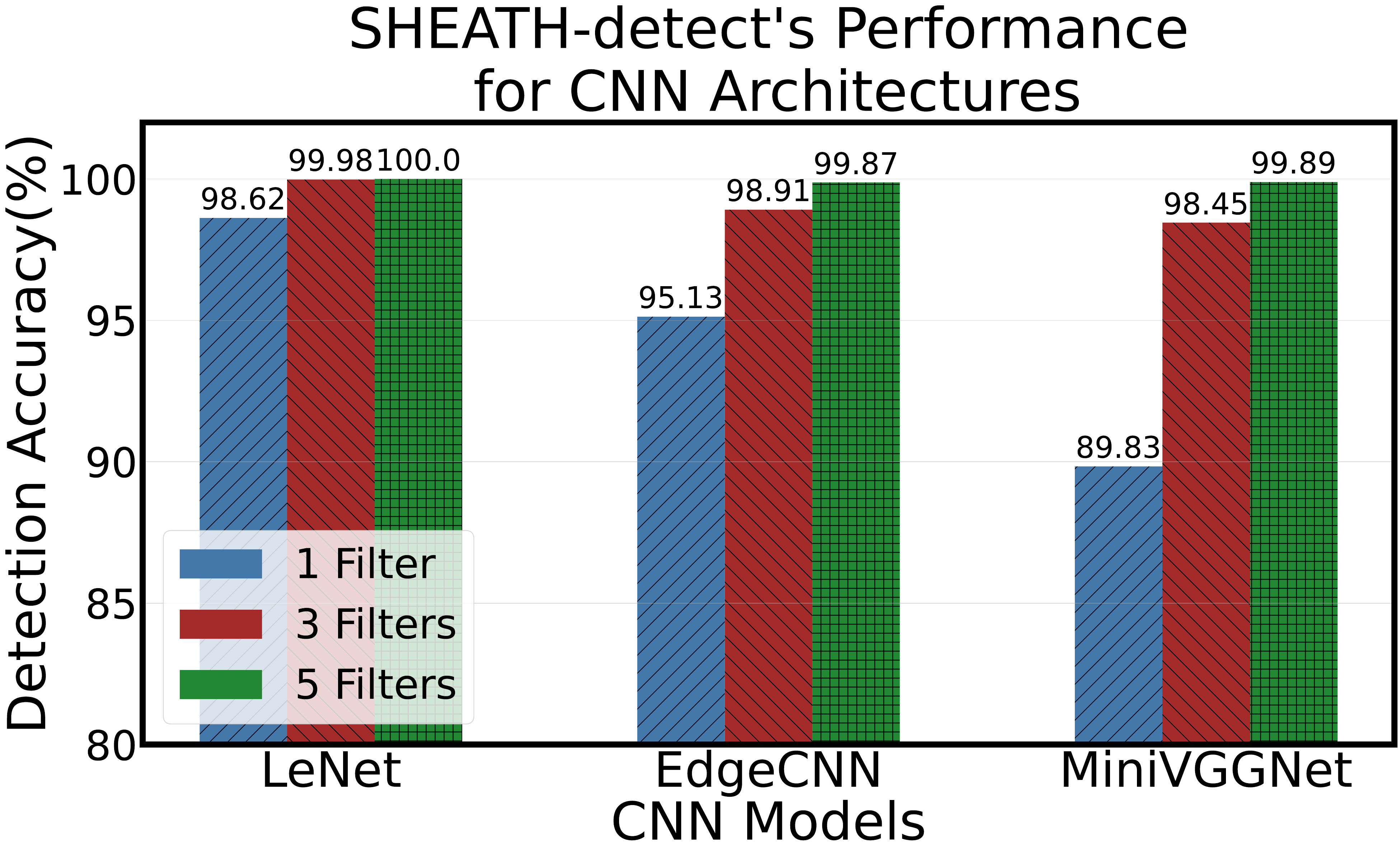}
    \vspace{-6pt}
    \caption{\small SHEATH's performance across different CNN architectures}
    \label{fig:rq6}
    \vspace{-9pt}
\end{figure}

\vspace{3pt}
\noindent\textbf{{\textit{RQ6 - SHEATH's Performance with Model Complexity (Scalability):}}} We explored how diverse CNN architectures influence performance to determine if SHEATH maintains efficacy with escalating model intricacy. The study included \textit{EdgeCNN}, LeNet, and MiniVGGNet. For \textit{EdgeCNN}, \textit{PseudoNet}'s accuracy increased from 95.13\% with one filter to 99.87\% with three. LeNet exhibited 98.62\% accuracy for one filter, achieving 100\% by three filters. MiniVGGNet began at 89.83\% with one filter, escalating to 99.89\% by the third. As illustrated in Fig.~\ref{fig:rq6}, these findings highlight SHEATH's adaptable and robust performance across various complexities.

\begin{figure}[t]
    \centering
    \includegraphics[width=0.8\columnwidth]{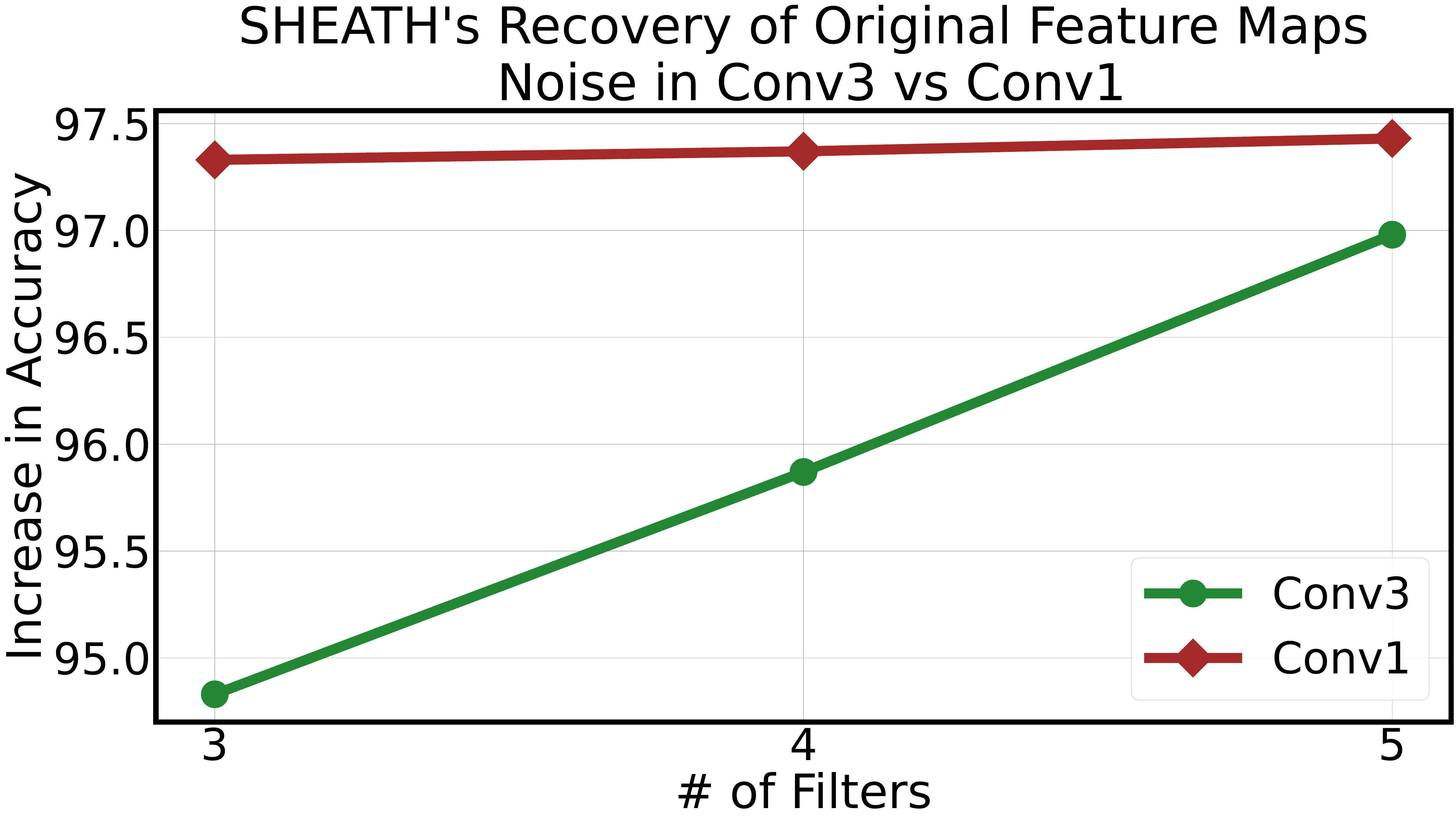}
    \vspace{-5pt}
    \caption{\small Impact of location of noise injection on SHEATH.}
    \label{fig:rq10}
    \vspace{-12pt}
\end{figure}

\vspace{-3pt}
\noindent\textbf{\textit{RQ7 - Impact of location of noise injection on SHEATH:}} Fig.~\ref{fig:rq10} shows the SHEATH-\textit{Recover} module's performance when the noise is injected in (a) earlier layer and (b) last layer of the model, \textit{``Conv1"} and \textit{``Conv3}, respectively with an increasing number of filters. As the number of filters increases, SHEATH's accuracy when  \textit{``Conv1} is under attack is better than when \textit{``Conv3} is under attack, emphasizing that SHEATH effectively detects adversarial noise and eliminates the noise impacts regardless of the location.

\begin{figure}[t]
    \centering
    \includegraphics[width=0.75\columnwidth]{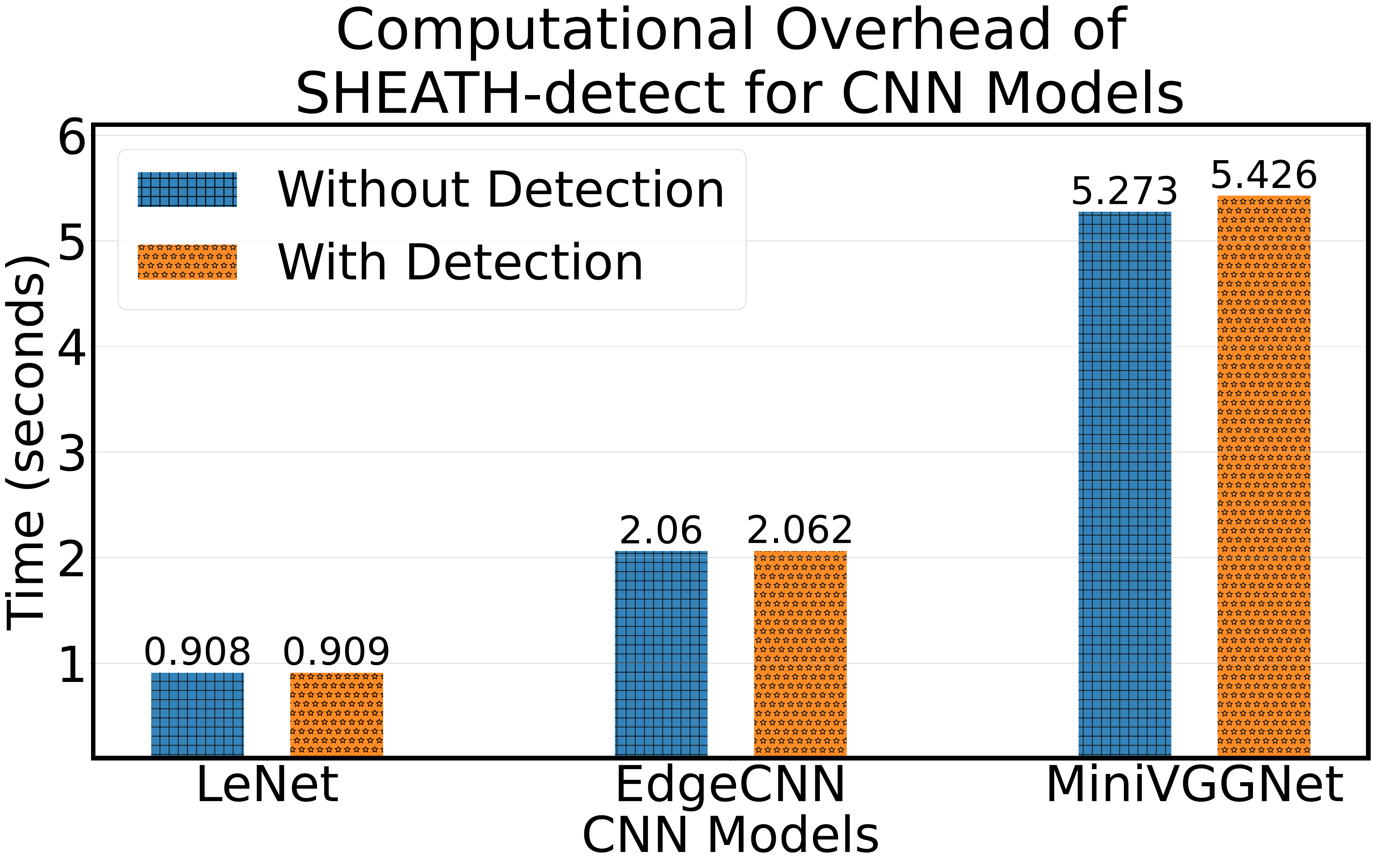}
    \vspace{-5pt}
    \caption{\small SHEATH-detect's computational overhead.}
    \label{fig:rq2}
    \vspace{-15pt}
\end{figure}

\vspace{3pt}
\noindent\textbf{\textit{RQ8 - Computational Overhead of SHEATH in real-time
CNN environments:}}  
%
When deploying verification tools like SHEATH in real-time CNN environments, assessing the computational overhead is essential to understand the extra processing time and resources. For optimal system performance, this should be minimal, even as it improves model robustness. We measured this overhead as the increase in processing time for models using SHEATH compared to their baseline computation time without it. Measurements were carried out on a system with a 12th Gen Intel(R) Core(TM) i5-1235U processor, 1.30 GHz frequency, 16.0 GB RAM, running a 64-bit operating system, assisted by a T4 GPU on Google Colab for consistent execution. In a practical setup, only the model training will be done on resources like Colab, whereas the inference will be computed on the edge nodes (IoT devices). LeNet had an increase of 1.7ms (0.908s to 0.9097s), \textit{EdgeCNN} an increase of 1.87ms (2.06s to 2.06187s), and MiniVGGNet, a more noticeable overhead of 153ms (5.273s to 5.426s). In summary, SHEATH-\textit{Detect} introduces a minimal computational overhead, as shown in Fig.~\ref{fig:rq2}, ensuring its potential suitability for real-time deployments. 

Furthermore, using empirical data, we evaluated the performance of SHEATH against the redundancy approach regarding time complexity and memory overhead. The \(t_{\text{d}}\), {\(t_{\text{m}}\)}, and \(t_{\text{r}}\) are 0.2033s, 0.4632s, and 0.2784s, respectively. The worst-case time complexity for SHEATH (when both detection and correction/recovery are applied) is $t_{{\text{d}}} + t_{\text{m}} = 0.6665$ seconds. However, during infrequent attacks or noise scenarios, SHEATH's average performance time is $t_{\text{d}}$, making it competitive against $t_{\text{r}}$. Similarly, \(m_{\text{d}}\), \(m_{\text{m}}\), and \(m_{\text{r}}\) are 0.0Mb, 0.015625Mb, and 0.0Mb, respectively. The memory overhead for the SHEATH-\textit{Recover} module is minimal, with an increased overhead of $m_{\text{m}}$ MiB compared to redundancy. While the worst-case time complexity of SHEATH is slightly higher than doing redundancy, its average performance, especially under infrequent attack scenarios, is competitive. With its adaptability, flexibility in handling threats, and minimal memory overhead, SHEATH is a justifiable and efficient alternative to redundancy. In case of frequent attacks, redundancy will be a better alternative; however, an HC environment with frequent attacks will not be favorable.  

\vspace{3pt}
\noindent\textbf{\textit{RQ9 - SHEATH's efficacy in case of multiple nodes:}} We implemented the deployment scenario as depicted in Fig.~\ref{fig:rq10}. In the \textit{EdgeCNN} model, the noise was added to the nodes having \textit{``Conv1"} and \textit{``Conv3"}. If the impact of noise in \textit{``Conv1"} is not mitigated, it will lead to a cascading effect, which will further decrease the model accuracy when \textit{``Conv3"} is also noisy. Hence, we deployed SHEATH for both layers/nodes and compared the model accuracy with and without SHEATH. As shown in Table.~\ref{tab:merged_accuracy_comparison}, SHEATH successfully recovers the original feature maps for both \textit{``Conv1"} and \textit{``Conv3"}. Hence, the model maintains a high accuracy consistently, regardless of the noise levels. It is worth mentioning that the accuracy is 97.55\% for all noise levels, as it is measured on the same dataset using the same recovery modules.


\vspace{3pt}
\noindent\textbf{\textit{RQ10 - SHEATH's efficacy in recovering original feature maps:}} Table~\ref{tab:merged_accuracy_comparison} provides a comprehensive evaluation of the SHEATH-Recover module's effectiveness in identifying and rectifying noise disturbances, particularly under Gaussian Noise Attack and Polarity Switch Attack scenarios. Utilizing the Pseudonet framework within the EdgeCNN architecture, the model generated five feature maps, while the SHEATH-Recover module effectively recovered the remaining 59 feature maps. Specifically, amidst Gaussian Noise perturbations, the model's accuracy improved significantly with SHEATH. At a standard deviation of 50\% and a noise level of 65\%, the model's accuracy improved from 18.43\% without SHEATH to 97.55\% with SHEATH. Notably, even under a higher noise intensity, such as at a standard deviation of 90\%, the model's accuracy showed improvement from 12.51\% to 97.55\%. 
Similarly, for polarity switch attacks, and at a noise level of 85\% the model's accuracy improved from 6.26\% without SHEATH to 97.36\% with SHEATH. Furthermore, at higher noise levels, such as at a noise level of 95\%, the accuracy improved from 4.23\% to 97.36\%. Hence, SHEATH can efficiently recover original feature maps amidst varying levels of noise injection.

\section{Related Works}
\label{sec:related_works}

We explore related works in terms of mechanisms of collaborative inference and security challenges in HC.
\subsection{Collaboration Inference}

Collaborative inference in edge devices, especially in IoT, has been a topic of significant interest. The drive to optimize the deployment of deep learning models across distributed edge devices has led to significant advancements in collaborative inference mechanisms. Huang et al.~\cite{huang2022decentralized} proposed a decentralized and collaborative deep learning inference system, DeColla, designed explicitly for intelligent IoT devices. This system migrates DNN computations to IoT devices, enhancing efficiency and robustness. Following this, Shlezinger et al.~\cite{shlezinger2021collaborative} introduced an edge ensemble framework for collaborative inference. Their approach aimed to leverage multiple edge devices to improve prediction accuracy, especially in environments with limited computational resources. Disabato et al.~\cite{disabato2021distributed} presented a methodology for distributing the computation of CNNs onto IoT units. Their approach focused on enabling real-time recall and autonomous decision-making, addressing the challenges of implementing deep learning solutions on resource-constrained IoT devices. Naveen et al.~\cite{naveen2021low} developed a model to optimize the allocation of deep learning tasks across edge nodes. Their primary goal was to minimize inference latency, ensuring faster and more efficient deep learning inference in distributed IoT edge clusters. Du et al.~\cite{du2020distributed} proposed a distributed in-situ CNN inference system tailored for IoT applications. Their system utilized a loosely coupled structure, synchronization-oriented partitioning, and decentralized asynchronous communication to address the resource constraints of individual IoT devices.



\begin{table}[t]
\centering
\caption{Comparison of Model Accuracies with and without SHEATH when subjected to Different Noise Types}
\begin{tabular}{|c|c|c|c|c|}
\hline
\textbf{Std Dev} & \textbf{Noise} & \multicolumn{2}{c|}{\textbf{Model Accuracy}} & \textbf{Noise} \\
\textbf{(\%)}    & \textbf{Level} & \textbf{with Noise} & \textbf{with SHEATH} & \textbf{Type}  \\
\hline
10 & 65 & 0.8549 & 0.9755 & Gaussian Noise \\
\cline{1-4}
50 & 65 & 0.1843 & 0.9755 & Gaussian Noise \\
\cline{1-4}
90 & 65 & 0.1251 & 0.9755 & Gaussian Noise \\
\hline
- & 75 & 0.1039 & 0.9736 & Polarity Switch \\
\cline{1-4}
- & 85 & 0.0626 & 0.9736 & Polarity Switch \\
\cline{1-4}
- & 95 & 0.0423 & 0.9736 & Polarity Switch \\
\hline
\end{tabular}
\label{tab:merged_accuracy_comparison}
\vspace{-6pt}
\end{table}

Furthermore, Hemmat et al.~\cite{hemmat2022edgen} introduced the EdgeAI framework, which focuses on distributed inference with local edge devices. Their approach aimed to achieve minimal latency using partitioning, pruning, and parallel inference techniques. Zhang et al.~\cite{zhang2021slicing} proposed DeepSlicing, a collaborative CNN inference system. By combining data and model partitioning, they aimed to enable efficient parallel inference with reduced latency, addressing the challenge of low-latency collaborative inference for CNNs on edge devices. Mao et al.~\cite{maomodnn} presented MoDNN, a locally distributed mobile computing system. Their system was designed to accelerate DNN computations on resource-constrained mobile devices by partitioning techniques and reducing non-parallel data delivery time. Zhao et al.~\cite{zhaodeepthings} introduced the DeepThings framework, which focuses on the deployment of CNNs on resource-constrained IoT edge clusters. Their approach utilized scalable partitioning, distributed work stealing, and optimized work scheduling to ensure efficient CNN-based inference applications. Zeng et al.~\cite{zeng2021coedge} proposed CoEdge, a distributed DNN computing system. Their system orchestrates cooperative DNN inference over heterogeneous edge devices with the primary goal of reducing overall latency by using adaptive workload partitioning and ensuring energy-efficient execution.

\subsection{Security in HC Inference}

The security of collaborative inference systems, especially in decentralized networks, is paramount. With the proliferation of edge devices and IoT, ensuring the integrity and confidentiality of data and models becomes a significant challenge. Odetola et al.~\cite{odetola2021feshi} introduced FeSHI, a stealthy hardware Trojan attack that targets CNN layer feature maps. By exploiting the statistical attributes of these feature maps, they were able to design stealthy triggers and payloads, achieving misclassification at the CNN output without full knowledge of the architecture. Mohammed et al.~\cite{mohammed2020secure} highlighted the potential security risks in distributed CNNs. They proposed Trojan attacks on CNNs implemented across multiple nodes, demonstrating the potential for misclassification in distributed CNN inference. Adeyemo et al.~\cite{adeymo2023stain} explored the vulnerabilities of DNN models to adversarial noises, such as the Fast Gradient Signed Method (FGSM) and Gaussian Noise Perturbation (GNP). Their research aimed to determine attack detection accuracy and devise strategies to bolster system security against adversarial attacks. Furthermore, Odetola et al. proposed the first hardware accelerator for adversarial attacks based on memristor crossbar arrays, aiming to significantly improve the throughput of a visual adversarial perturbation system, which can further enhance the robustness and security of future deep learning systems~\cite{odetola2022hardening}. Baccour et al.~\cite{hamdi2020distprivacy} introduced the DistPrivacy methodology to secure sensitive data on DNNs. Their approach balanced latency, privacy level, and limited resources, ensuring that sensitive data remains confidential even in a distributed inference environment. Despite these works, there remains a gap in detecting adversarial noise-based attacks on HC-based CNNs and developing countermeasures. Our work with SHEATH aims to bridge this gap by both detecting the noise and eliminating its impacts by recovering the original feature maps in case of detection. 

\section{Conclusion}
\label{sec:conclusion}

In conclusion, we proposed a novel framework, SHEATH, to secure HC-based CNN architectures against adversarial noise. 
SHEATH, first, detects the adversarial noise and, then, eliminates its effect on CNN inference by recovering
the original feature maps in case of a detected attack. It operates without requiring complete knowledge of the CNN model, ensuring robust data and model protection. 
We tested and validated SHEATH's performance under diverse attack scenarios. Our experiments reiterate SHEATH's adaptability and effectiveness across different CNNs to maintain a high inference accuracy even under attacks. For instance, CNN's overall model accuracy was increased from 18.43\% without SHEATH to 97.55\% with SHEATH despite a noise injection attack while incurring minimal overhead.
In the future, we will extend the proposed concept for collaborative infrastructures without trusted nodes and test SHEATH on an FPGA-based hardware HC implementation to further reduce the overhead.

\section*{Acknowledgement}

This research was supported in part by the Air Force Office of Scientific Research (AFOSR)/ Air Force Research Laboratory (AFRL) Summer Faculty Fellowship Program (SFFP) for summer 2023 and the Department of Energy (DOE) under Award DE-NA0004016. 
The views and conclusions contained herein are those of the authors and should not be interpreted as necessarily representing the official policies or endorsements, either expressed or implied, of the AFRL, DOE, or the U.S. Government.



 
%

\bibliographystyle{unsrt}
\bibliography{References}


 





\end{document}